\documentclass[final,3p,times,sort&compress]{elsarticle}
\usepackage{amssymb}
\usepackage{amsthm}
\usepackage{lineno}
\modulolinenumbers[5]

\DeclareMathAlphabet{\mathcal}{OMS}{cmsy}{m}{n}
\usepackage[framed,numbered,autolinebreaks,useliterate]{mcode}
\usepackage[hyphens]{url}
\usepackage[hidelinks,breaklinks]{hyperref}
\usepackage[version=3,arrows=pgf-filled]{mhchem} 
\usepackage{bm} 
\usepackage{algpseudocode}
\usepackage{algorithm}
\usepackage{color,colortbl,booktabs,array}
\usepackage{multirow}

\usepackage{xr}

\algrenewcommand\textproc{} 
\algdef{SE}[SUBALG]{Indent}{EndIndent}{}{\algorithmicend\ }%
\algtext*{Indent}
\algtext*{EndIndent}

\algnewcommand\algorithmicswitch{\textbf{switch}}
\algnewcommand\algorithmiccase{\textbf{case}}
\algnewcommand\algorithmicassert{\texttt{assert}}
\algnewcommand\Assert[1]{\State \algorithmicassert(#1)}%

\algdef{SE}[SWITCH]{Switch}{EndSwitch}[1]{\algorithmicswitch\ #1\ \algorithmicdo}{\algorithmicend\ \algorithmicswitch}%
\algdef{SE}[CASE]{Case}{EndCase}[1]{\algorithmiccase\ #1}{\algorithmicend\ \algorithmiccase}%
\algtext*{EndSwitch}%
\algtext*{EndCase}%
\definecolor{applegreen}{rgb}{0.55, 0.71, 0.0}
\definecolor{myred}{RGB}{162, 20, 47}
\definecolor{myblue}{RGB}{44, 137, 160}
\definecolor{OrangeInks}{rgb}{1.0, 0.67, 0}
\definecolor{BlueInks}{rgb}{0, 0.67, 1.0}
\definecolor{GreenInks}{rgb}{0.31, 0.6, 0}
\definecolor{BlackInks}{rgb}{0, 0, 0}
\definecolor{WhiteInks}{rgb}{1, 1, 1}
\definecolor{RedInks}{rgb}{1, 0, 0}
\definecolor{BlueInks2}{rgb}{0, 0.5, 0.8}
\definecolor{darkblue}{RGB}{33, 103, 120}
\definecolor{lightblue}{RGB}{55, 171, 200}
\definecolor{lightgray}{RGB}{111, 145, 138}

\newcommand{\rev}[1]{{\color{black}{#1}}}
\newcommand{\new}[1]{{\color{black}{#1}}}

\newcommand{\setcolordiamonds}[1]{{\color{darkblue}{#1}}}
\newcommand{\setcolorcircles}[1]{{\color{lightblue}{#1}}}
\newcommand{\setcolortriangles}[1]{{\color{lightgray}{#1}}}
\newcommand{\setcolorsquares}[1]{{\color{myred}{#1}}}
\def\D{\mathrm{d}}
\newcommand{\blackline}{\raisebox{2pt}{\tikz{\draw[-,BlackInks,solid,line width = 0.9pt](0,0) -- (5mm,0);}}}
\newcommand{\blueline}{\raisebox{2pt}{\tikz{\draw[-,myblue,solid,line width = 0.9pt](0,0) -- (5mm,0);}}}

\begin{document}

\begin{frontmatter}

\title{Combustion Toolbox: An open-source thermochemical code for gas- and condensed-phase problems involving chemical equilibrium}

\author{Alberto Cuadra\corref{author}}
\author{César Huete}
\author{Marcos Vera}

\cortext[author] {Corresponding author.\\\textit{E-mail address:} acuadra@ing.uc3m.es (A. Cuadra)}
\address{Departamento de Ingenier\'{\i}a T\'{e}rmica y de Fluidos, Escuela~Polit\'{e}cnica~Superior, Universidad~Carlos~III~de~Madrid, 28911 Legan\'{e}s, Spain}

\begin{abstract}

The Combustion Toolbox (CT) is a newly developed open-source thermochemical code designed to solve problems involving chemical equilibrium for both gas- and condensed-phase species. The kernel of the code is based on the theoretical framework set forth by NASA's computer program CEA (Chemical Equilibrium with Applications) while incorporating new algorithms that significantly \rev{improve both convergence rate and robustness}. The thermochemical properties are computed under the ideal gas approximation using an up-to-date version of NASA's 9-coefficient polynomial fits. These fits use the Third Millennium database, which includes the available values from Active Thermochemical Tables. Combustion Toolbox is programmed in MATLAB with \rev{an object-oriented} architecture composed of three main modules: CT-EQUIL, CT-SD, and CT-ROCKET. The \new{kernel} module, CT-EQUIL, minimizes the Gibbs/Helmholtz free energy of the system using the technique of Lagrange multipliers combined with a multidimensional Newton-Raphson method, upon the condition that two state functions are used to define the mixture properties (e.g., enthalpy and pressure). CT-SD solves processes involving strong changes in dynamic pressure, such as steady shock and detonation waves under normal and oblique incidence angles. Finally, CT-ROCKET estimates rocket engine performance under highly idealized conditions. The new tool is equipped with a versatile Graphical User Interface and has been successfully used for teaching and research activities over the last four years. Results are in excellent agreement with CEA, Cantera within Caltech's Shock and Detonation Toolbox (SD-Toolbox), and the Thermochemical Equilibrium Abundances (TEA) code. CT is available under an open-source GPLv3 license via GitHub \new{\url{https://github.com/CombustionToolbox/combustion_toolbox}}, and its documentation can be found in \url{https://combustion-toolbox-website.readthedocs.io}.

\end{abstract}


\end{frontmatter}

\section{Introduction}

The computation of chemical equilibrium has been widely used during the last century to determine the composition of multi-component mixtures subject to complex thermochemical transformations. The resulting mathematical problem is simple in systems involving only a few species, such as the complete combustion of rich hydrocarbon-air mixtures, or \new{the} dissociation of diatomic gas mixtures, e.g., air at moderate temperatures. However, the incomplete combustion of a typical hydrocarbon, for instance, methane, with air, involves hundreds of reactions and more than fifty species~\cite{smith1999}\new{, which} makes finding the final equilibrium state of the products a challenging task. 

Two equivalent methods can be employed to determine the composition of the products at equilibrium: using equilibrium constants or minimizing the Gibbs/Helmholtz free energy of the system~\cite{zeleznik1960}. The first method requires specifying a sufficiently large set of elementary reactions at equilibrium (see, e.g., Refs.~\cite{brinkley1947, paz2013, leal2015, woitke2018, stock2018, stock2022, kitzmann2024}). This favors the second, where each species is treated independently, and the focus is shifted to the chemical potentials of the different species involved~\cite{van1970, smith1982}. The second method was first introduced by the pioneering work of White in 1958~\cite{white1958} and has become the cornerstone in the development of virtually all state-of-the-art thermochemical codes~\cite{zeleznik1968, eriksson1971, reynolds1986, michelsen1989, gordon1994, vovnka1995, eriksson2004, nichita2002, pope2003, pope2004, neron2012, scoggins2015, blecic2016, gray2017, tsanas2017a, tsanas2017b, coatleven2022, combustiontoolbox}.

The solution to the resulting minimization problem requires the evaluation of the thermodynamic properties of all species involved at a given temperature. For this purpose, extensive thermodynamic databases have been compiled, such as the NIST-JANAF tables~\cite{chase1998, dorofeeva2001}, NASA's polynomials~\cite{Mcbride2002} and, more recently, the Third Millennium (Burcat) Database~\cite{burcat2005} with updates from Active Thermochemical Tables (ATcT)~\cite{ruscic2005} to evaluate the enthalpies of formation. The ATcT rely on the use of a complete thermochemical network (TN) instead of the more traditional datasets based on individual reactions. The use of the full TN yields more accurate results that are, in addition, fully documented (uncertainties included). But more important is the fact that databases are easily updated with new knowledge, and readily provide new values for the thermochemical properties of all species~\cite{ruscic2019_book}.

A recent work by Scoggins et al.~\cite{scoggins2017} reported an exhaustive review of over 1200 unique chemical species from several databases that fully (Goldsmith~\cite{goldsmith2012} and Blanquart~\cite{blanquart2007, blanquart2009, narayanaswamy2010, blanquart2015}), partially (Burcat), or did not (NASA) rely on the ATcT. They identified significant differences between Burcat's and NASA's databases due to the inconsistency of the species enthalpy of formation in the latter. Although this type of analysis is out of the scope of this work, all chemical species from Burcat's database are also available in Combustion Toolbox (CT) and can be identified by the suffix ``\_M''. Thus, when both databases contain a given species, the final choice is left to the user. 

Thermochemistry is firmly rooted in the study of combustion problems, high-speed flows, reactive and non-reactive shocks, rocket engine performance, and high explosives~\cite{miller1990, kubota2015}. For instance, strong hypersonic shocks involve changes in the molecular structure of the gas, including vibrational excitation leading to dissociation~\cite{huete2021, jiang2022}, and later electronic excitation leading to ionization~\cite{candler1991}, which eventually transform the gas into a plasma. Turbulent combustion and gaseous detonations have also been the topic of intense research due to their high thermodynamic efficiency in propulsion applications ~\cite{Libby94, wolanski2013}. But they often exhibit strong deviations from equilibrium due to the wide range of length and time scales involved, making it necessary to rely on complex fluid dynamical analyses and numerical simulations with a high computational cost~\cite{Di2021, raman2022}. Despite the deep understanding provided by the latter approach, there are still cases in which a proper physical explanation can not be found based only on numerical results. In these cases, separation of scales may allow to split the problem into simpler ones, where the assumption of chemical equilibrium could be justified in some representative scenarios~\cite{Cuadra2020, huete2021, cuadra2023_aiaa}.

Chemical equilibrium can be formulated either for single-phase gas mixtures~\cite{white1958, pope2003, pope2004, eriksson2004, blecic2016, gray2017, stock2018, stock2022}, single-phase gas mixtures with pure-condensed species---as in NASA's CEA code \cite{gordon1994}, Refs.~\cite{brinkley1947, eriksson1971, woitke2018}, and this work \cite{combustiontoolbox}---or for multi-phase systems~\cite{reynolds1986, michelsen1989, michelsen1994, vovnka1995, nichita2002, neron2012, paz2013, scoggins2015, leal2015, tsanas2017a, tsanas2017b, scoggins2020, coatleven2022, kitzmann2024}. The latter case requires special treatments to ensure that the global \new{minimum is} reached, due to the non-convexity of the Gibbs free energy \cite{nichita2002, neron2012, leal2017}. These include global optimization methods like the tunneling method~\cite{groenen1996} or differential evolution~\cite{storn1997, price2013}.

The above review has identified many thermochemical codes currently available to the community. Nevertheless, to the best of our knowledge, there is not yet an open-source code based on up-to-date databases, written in a high-level programming language, fully documented, with high-performance computing capabilities, able to model a wide variety of applications, and equipped with a user-friendly Graphic User Interface (GUI). Combustion Toolbox was conceived with these long-term goals in mind, and is now presented and validated in this work. This MATLAB-GUI thermochemical code represents the core of an ongoing research work \new{and} has been used to investigate a series of problems during the last few years \new{\cite{Cuadra2020, huete2021, simex2022, cuadra2023_aiaa, cuadra2023_thesis, dahake2023, chen2024, xu2024, qin2024}}. Results are in excellent agreement with NASA's CEA code~\cite{gordon1994}, Cantera~\cite{Cantera} within Caltech's Shock and Detonation Toolbox (SD-Toolbox)~\cite{Browne2008SDT, Browne2008}, and the Thermochemical Equilibrium Abundances (TEA) code~\cite{blecic2016}.

The paper is structured as follows. Section~\ref{sec:overview} starts with an initial overview of CT. The equilibrium kernel (CT-EQUIL) is presented in Section~\ref{sec:EQUIL}. The shock and detonation module (CT-SD) is discussed in Section~\ref{sec:SD}, followed by the rocket performance module (CT-ROCKET) in Section~\ref{sec:ROCKET}. The results of all modules are validated against other codes in Sections~\ref{sec:EQUIL} to \ref{sec:ROCKET}. A detailed description of the GUI is given in Section~\ref{sec:GUI}. And, finally, the conclusions are presented in Section~\ref{sec:conclusions}.

\section{Overview of Combustion Toolbox}
\label{sec:overview}
Combustion Toolbox~\cite{combustiontoolbox} is a GUI-based thermochemical code written in MATLAB with an equilibrium kernel based on the mathematical formulation set forth by NASA's CEA code~\cite{gordon1994}. The thermodynamic properties of the gaseous species are modeled with the ideal gas equation of state (EoS), and an up-to-date version of NASA's 9-coefficient polynomial fits \new{(NASA9) based on}~\cite{Mcbride2002,burcat2005,ruscic2005}. CT is a new thermochemical code written from scratch in a \rev{object-oriented} architectural format composed of three main modules: CT-EQUIL, CT-SD, and CT-ROCKET. 

The \new{kernel} module, CT-EQUIL, computes the composition at the equilibrium of multi-component gas mixtures that undergo canonical thermochemical transformations from an initial state (reactants), defined by its initial composition, temperature, and pressure, to a final state (products), defined by a set of chemical species (in gaseous---included ions---or pure condensed phase) and two thermodynamic state functions, such as enthalpy and pressure, e.g., for isobaric combustion processes. CT-SD solves steady-state shock and detonation waves in either normal or oblique incidence. Finally, CT-ROCKET computes the theoretical performance of rocket engines under highly idealized conditions. Even though all modules are enclosed in a user-friendly GUI, they can also be accessed from MATLAB's command line in plain code mode. 

There is a fourth closed-source (i.e., proprietary) module, CT-EXPLO, that estimates the theoretical properties of high explosive mixtures and multi-component propellants with non-ideal EoS. Although still under development, CT-EXPLO is distributed in its current form as the thermochemical module of SimEx~\cite{simex2022} subject to a proprietary license. Further details on this module will be provided elsewhere.

\subsection{Software Architecture}

As previously mentioned, the program was developed from scratch using MATLAB and \rev{an object-oriented} architectural design. All the modules rely on CT-EQUIL to compute the thermodynamic properties of the species involved and the chemical composition of the mixtures at equilibrium. The full package can be accessed from the GUI (see Section~\ref{sec:GUI}) or \new{directly} from MATLAB's command line. Additionally, the main computations performed using the GUI are callbacks to the plain code. Consequently, any change made to the code is immediately reflected in the GUI, leading to a more flexible and adaptable tool.

MATLAB was selected as programming language due to its excellent linear algebra, visualizing and debugging capabilities, extensive documentation, active community, and dedicated app development framework (App Designer). \rev{CT has significantly evolved since its version 1.0.5~\cite{combustiontoolboxv1, cuadra2023_thesis}, which relied solely on procedural techniques for performance advantages over MATLAB Object-Oriented Programming (OOP). However, the current version, v1.1.0~\cite{combustiontoolbox}, has been completely rewritten using the OOP paradigm due to the inherent limitations of procedural techniques, such as reduced code modularity and scalability. After meticulous optimization, the new object-oriented-based code has achieved substantial performance improvements, achieving between $1.3\times$ and $1.9\times$ better performance compared to the previous version under identical test conditions.} Even though MATLAB is an interpreted language, which introduces a significant performance cost compared to other (compiled) languages \cite{andrews2012}, CT computational times are still competitive compared to similar codes. For instance, as shown in Section~\ref{sec:SD}, CT is about one order of magnitude faster than the MATLAB version of Caltech's SD-Toolbox used with Cantera (written in C++). However, there is room for further improvement by isolating specific demanding tasks into C++ subroutines and accessing them from MATLAB using MEX files. \rev{Moreover, the adoption of OOP in the current version facilitates the parallelization of the code. These optimization techniques will be explored in future code releases.}

\rev{The classes and methods included in CT's modules (described in detail later) rely on several \textit{core} and \textit{database} classes. \rev{The most relevant among these are:}
\begin{itemize}
    \item \texttt{Species}: stores the properties and thermodynamic functions of a chemical species.
    \item \texttt{ChemicalSystem}: encapsulates the chemical species and phases within a chemical system.
    \item \texttt{Mixture}: represents a multi-component mixture, including its thermodynamic properties and the solution of the problem at hand.
    \item \texttt{EquationState}: defines the thermodynamic equation of state for a \texttt{Mixture}.
    \item \texttt{Database}: an abstract superclass with common methods and properties for database-related classes.
    \item \texttt{NasaDatabase}: constructs NASA's thermochemical database \cite{Mcbride2002} using a structure of \texttt{Species} objects. By default, this class also includes data from Burcat's database \cite{burcat2005}, obtained in its NASA9 format.
\end{itemize}
}

For efficient data access, \rev{the \texttt{Species} class} employs \textit{griddedInterpolant} objects (see MATLAB's built-in function \texttt{griddedInterpolant.m}), which use piecewise cubic Hermite interpolating polynomials (PCHIP)~\cite{fritsch1980}. This approach speeds up the \new{thermodynamic} data access by a factor of 200\% with respect to the evaluation of NASA's polynomials. \new{Starting from v1.1.0, these objects are also temporally allocated in cache for even better performance.} Furthermore, for temperatures outside the bounds, we avoid the higher order terms of the polynomials by linear extrapolation, similar to Ref.~\cite{stock2018}, extending the range of validity of the thermodynamic data available. It should be emphasized that this extension is limited to a narrow temperature range and may not apply to temperatures significantly outside \new{the bounds}.

\rev{The Combustion Toolbox is organized using namespaces. Hence, it is necessary to import the different subpackages into MATLAB workspace as follows}
\vspace{-0.3cm}

\begin{minipage}{0.9\linewidth}
\begin{lstlisting}[basicstyle=\footnotesize]
import combustiontoolbox.databases.NasaDatabase
import combustiontoolbox.core.*
\end{lstlisting}
\end{minipage}

\rev{\noindent The first statement imports a specific class or method, while the second imports an entire subpackage. To start working with CT, we also need to load/generate a database and initialize the chemical system.} This can be easily done with one of the following sample statements:
\vspace{-0.3cm}

\begin{minipage}{0.9\linewidth}
\begin{lstlisting}[basicstyle=\footnotesize]
system = ChemicalSystem(DB)
system = ChemicalSystem(DB, 'Soot formation extended')
system = ChemicalSystem(DB, {'N2', 'O2', 'NO', 'N', 'O'})
\end{lstlisting}
\end{minipage}

\rev{\noindent In these examples, \texttt{DB} represents an instance of the database class used, e.g., \texttt{DB = NasaDatabase()}. The first option initializes the system with all the possible species that could arise based on the elements present in the reactants (see class's method \texttt{findProducts}). The second option initializes the system with an extended list of 94 species that typically appear in CHON mixtures, including solid carbon C$_{(\rm gr)}$. Finally, the third option restricts the system to a specific list of species that may appear as products.}

\begin{figure}[!htpb]
    \centering
    \includegraphics[width=0.98\textwidth]{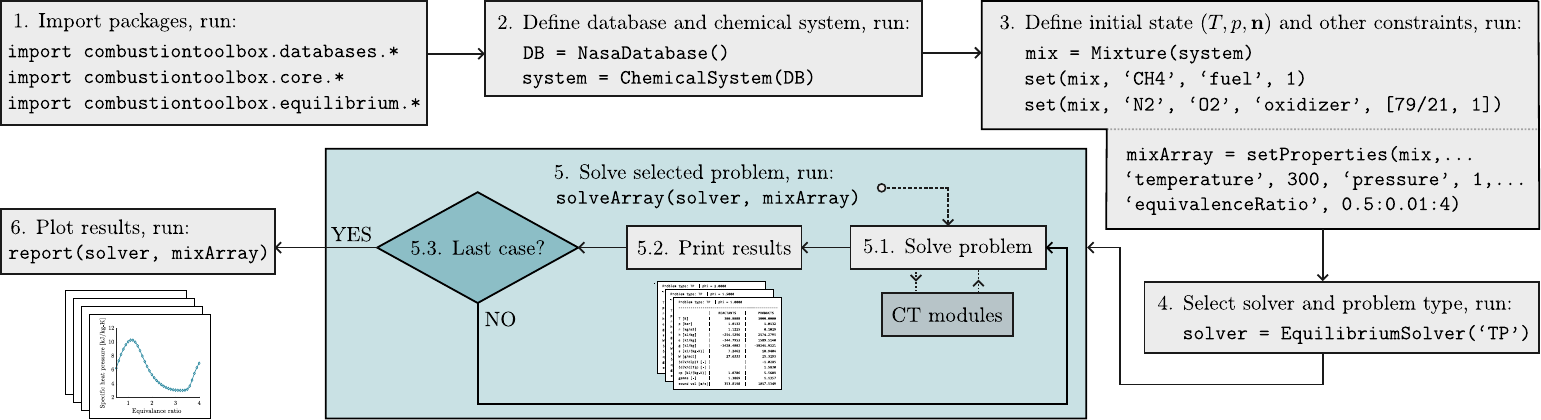}
    \caption{\rev{Combustion Toolbox simplified workflow.}}
    \label{fig:workflow}
\end{figure}

\rev{Figure~\ref{fig:workflow} summarizes the main steps required to solve a problem with CT. First, import the necessary packages. Second, define the database and initialize the chemical system using one of the statements indicated above. Then, specify the initial state of the mixture (temperature, pressure, and molar composition), select the solver, and define the problem type. Depending on the problem's constraints, additional parameters may be required, e.g., for planar shocks, it is necessary to specify the pre-shock velocity in m/s or, equivalently, the pre-shock Mach number. Once the setup is complete, invoke the \texttt{solveArray} method from the selected solver to compute the results. Finally, for post-processing, CT solvers include a built-in \texttt{report} method that shows predefined plots. Results can be exported using the \texttt{Export} class, which allows to save data into a spreadsheet or as a \texttt{.mat} file.}

The \new{CT source code} is organized into several top-layer folders: \rev{\textit{+combustiontoolbox}}, \textit{databases}, \textit{examples}, \textit{gui}, \textit{installer}, \textit{modules}, \textit{utils}, and \textit{validations}. \rev{The \textit{+combustiontoolbox} folder encapsulates the different modules (namespaces) implemented in CT, such as CT-EQUIL (\textit{+equilibrium}), CT-SD (\textit{+shockdetonation}), and CT-ROCKET (\textit{+rocket}).} The \textit{databases} folder mainly consists of raw data and \texttt{.mat} files that contain the thermochemical properties of the individual chemical species~\cite{Mcbride2002,burcat2005,ruscic2005}. The \textit{examples} folder includes various examples that demonstrate the wide variety of problems that can be solved with CT. The \textit{gui} folder contains the routines that are specifically designed for the GUI. The \textit{installer} folder contains all the installation files of the GUI: the MATLAB toolbox and the royalty-free stand-alone versions for different operative \new{systems}. This step is straightforward (see \texttt{INSTALL.m}), \new{and further details can be found on the} CT website (\url{https://combustion-toolbox-website.readthedocs.io/en/latest/install.html}). Finally, the \textit{validation} folder includes the routines used to validate CT with the results obtained with other codes, the unit testing files to ensure the correct functionality of the code, and all the graphs generated from these verifications.

\subsection{Collaborative framework and version control system}

The collaborative process of open-source code can greatly benefit from the contributions of other authors. However, in the absence of adequate tools, this process may become disorderly. To mitigate this risk, we employ Git as version control system (VCS) and GitHub as online hosting service. These technologies enable comprehensive tracking of all changes made to the code while maintaining complete transparency throughout the development process \cite{ram2013, chacon2014, blischak2016, perez2016}. To further ensure the integrity of the package, all contributions are subject to rigorous testing \new{before being merged} into the two primary branches (\texttt{master} and \texttt{develop}) using GitHub Actions, \new{providing} a powerful tool to retrace errors in the code.

\subsection{Documentation}

A notable fraction of open-source codes lack sufficient documentation, which impedes the code's usability and accessibility. To amend this issue, we use Sphinx~\cite{brandl2010}, a documentation generator written and used by the Python community, along with its MATLAB-domain extension~\cite{cederberg2023}. All function headers are written following Google's Python-style docstrings. The online documentation is hosted on \href{https://combustion-toolbox-website.readthedocs.io}{Read the Docs} and is regularly updated from the \href{https://github.com/CombustionToolbox/combustion_toolbox_website}{GitHub repository}. The new routines are automatically included in the online documentation using GitHub Actions. The framework allows having specific documentation for each distributed version. Additionally, the package includes several examples and all the validations carried out with other codes.

\subsection{Benchmarks}

The calculations presented in this work were performed on a laptop computer with the following specifications: Intel(R) Core(TM) i7-11800H CPU @ 2.30GHz with 8 physical cores and 64GB of RAM, running on a 64-bit Windows 11 Pro system and using MATLAB R2022b. The reported computation times represent the total elapsed time from the initialization of CT to the completion of all calculations. As a reference, loading the databases (approximately $3600$ species with their \textit{griddedInterpolant} objects) takes an average of \rev{0.7228} seconds, \new{which was} included in the reported times. \rev{Unless otherwise specified, the tolerance for the molar composition was consistently set to $10^{-14}$ throughout the study.}

\section{Thermochemical equilibrium module}
\label{sec:EQUIL}

This section presents the thermochemical equilibrium module\new{, CT-EQUIL, the kernel of the code}\rev{, encapsulated within the \textit{+equilibrium} subpackage}. This module \rev{relies on the \texttt{EquilibriumSolver} class, which} is composed of four main \rev{methods}: \rev{\texttt{equilibriumGibbs}}, \rev{\texttt{equilibriumHelmholtz}}, \rev{\texttt{equilibrateT}}, and \texttt{equilibrate}.

The first two, described in Sections~\ref{sec:EQUIL_kernel_1} and \ref{sec:EQUIL_kernel_2}, are used to compute the chemical equilibrium composition for given temperature-pressure (TP), or temperature-volume (TV) states, respectively. The third is employed to compute the thermodynamic properties of the mixture and is an upper layer of the previous routines. The \new{last one}, described in Section~\ref{sec:EQUIL_remainder}, represents the top layer of CT-EQUIL and is implemented to compute the chemical equilibrium composition and thermodynamic properties for any of the following pairs of specified state functions: TP, HP, SP, TV, EV, and SV, where T stands for temperature, P for pressure, H for enthalpy, S for entropy, E for internal energy, and V for volume.

Combustion Toolbox enables the computation of chemical equilibrium under various \new{caloric models} regarding the final gas mixture, including calorically perfect gas, calorically imperfect gas with frozen chemistry, or calorically imperfect gas with equilibrium chemistry, including dissociation and ionization. An example of these calculations is presented in Section~\ref{sec:SD}. It is also possible to freeze the composition of a subset of the species at equilibrium by defining them as inert species.

\subsection{Equilibrium composition at specified temperature and pressure (TP)}
\label{sec:EQUIL_kernel_1}

The \rev{\texttt{equilibriumGibbs}} routine computes the molar equilibrium composition $\mathbf{n} = \{n_1,n_2,\dots,n_{\rm NS}\}$ of a mixture $\mathbf{S} = \{S_1,S_2,\dots,S_{\rm NS}\}$ of NS species at a given temperature $T$ and pressure $p$ for a closed system by minimizing the Gibbs free energy $G\left(T,p,\mathbf{n}\right)$. This is an equality-constrained problem (ECP) subject to mass conservation, namely
\begin{subequations}
    \begin{gather}
    \min \, G(T, p, \mathbf{n}) \; \Leftrightarrow\; \D G(T, p, \mathbf{n}) = \sum\limits_{j\in \mathbf{S}}\mu_{j}(T, p,\mathbf{n}) \, \D n_j = 0, \label{eq:min_gibbs_a}\\ 
    q_i := \sum\limits_{j\in \mathbf{S}} a_{ij} n_j - b_i^\circ = 0,\quad \forall i \in \mathbf{E}
    \end{gather}
    \label{eq:chem_problem}%
\end{subequations}
where $n_j$ and $\mu_j$ are the number of moles and chemical potential of species $j$, respectively, $a_{ij}$ are the stoichiometric coefficients, i.e., the number of atoms of element $i$ per molecule of species $j$, and $b_i^\circ$ is the number of \rev{mols} of the $i$-th element in the initial mixture. There are as many linear constraints $q_i$ as NE elements $\mathbf{E} = \{E_1,E_2,\dots,E_{\rm NE}\}$ involved. The NS species can be either gaseous or condensed, assuming pure components. Thus, there are NG gaseous species $\mathbf{S^G} = \{S_1,S_2,\dots,S_{\rm NG}\}$ and $\rev{\rm{NC} = \ }\rm{NS} - \rm{NG}$ condensed species $\mathbf{S^C} = \{S_{\rm NG + 1},S_{\rm NG + 2},\dots,S_{\rm NS}\}$, where $\mathbf{S^G}$ and $\mathbf{S^C}$ are the respective subsets of $\mathbf{S}$. \rev{In addition, the system must comply with the Gibbs phase rule~\cite{smith2018}, ensuring that the number of elements does not exceed the number of chemical species present. This condition can be expressed as $\rm{NC}_s + 1 \leq \rm{NE} \leq \rm{NS}$, where NC$_s$ denotes the number of condensed species that remain stable while coexisting with the gas phase at chemical equilibrium.} Equation \eqref{eq:min_gibbs_a} must be supplemented with an EoS to define the thermodynamic functions. Our code implements the ideal gas EoS for the chemical potential
\begin{equation}
\mu_j\left(T,p,\mathbf{n}\right)=\mu_j^\circ\left(T\right)+\kappa_j RT\left(\ln{n_j} - \ln{\sum\limits_{j\in\mathbf{S^G}} n_j} +\ln{\frac{p}{p^\circ}}\right), \quad \forall j \in \mathbf{S}
    \label{eq:chemical_potential}
\end{equation}
where $\mu_j^\circ(T)$ is the chemical potential of species $j$ at the reference pressure ($p^\circ = 1$~bar) and the specified temperature, $\kappa_j$ is either one or zero depending on whether the species is in gaseous or condensed phase, and $R$ is the universal gas constant. As stated above, the CT-EQUIL module computes all thermochemical properties within the ideal gas approximation using an up-to-date version of NASA's 9-coefficient polynomial fits~\cite{Mcbride2002} that incorporates the Third Millennium database~\cite{burcat2005}, including the available values from Active Thermochemical Tables. For more details on the calculation of the thermochemical properties from the values contained in these databases\rev{, the reader is referred to the original sources~\cite{Mcbride2002, burcat2005}}.

\rev{The Gibbs phase rule constrains the number of condensed species that can appear at equilibrium, but it does not specify which of these NC species will be stable~\cite{kitzmann2024}. To determine the stability of the condensed phases, we directly impose the non-negative condition $n_j > 0, \forall j \in \mathbf{S^C}$ following the approach of Kulik et al.~\cite{kulik2013} and Leal et al.~\cite{leal2016,leal2017}. These inequalities are reformulated as a set of equality constraints by introducing slack variables $\varphi$, resulting in
\begin{equation}
    r_j := n_j \varphi_j - \tau = 0, \quad \forall j \in \mathbf{S^C}
    \label{eq:chem_problem_condensed}
\end{equation}
where we have included a constant $\tau = 10^{-25} \min(\mathbf{b}^\circ)$ for numerical stability purposes~\cite{leal2016}. In Eq.~\eqref{eq:chem_problem_condensed}, a condensed species is considered stable when $\varphi_j \approx 0$ and $n_j > 0$, whereas unstable condensed species is characterized by $\varphi_j > 0$ and $n_j \approx 0$.}

The ECP formulated in~\eqref{eq:chem_problem}-\eqref{eq:chem_problem_condensed} is solved using the method of Lagrange multipliers (an extended description thereof can be found in \cite{bertsekas2014}), which introduces the Lagrangian function
\begin{equation}
    \boldsymbol{\mathcal{L}}(T, p, \mathbf{n}, \boldsymbol{\lambda}, \boldsymbol{\varphi}) = G(T, p, \mathbf{n}) + \boldsymbol{\lambda}\, \mathbf{q}(\mathbf{n}) \rev{\ -\ \boldsymbol{\varphi}\, \mathbf{r}(\mathbf{n})},
     \label{eq:lagrangian}
\end{equation}
where $\boldsymbol{\lambda}$ \rev{and $\boldsymbol{\varphi}$ represent the multiplier vectors, of length NE and NC, respectively}. If there is an infinitesimal change, $\D\mathbf{n}$ and $\D\boldsymbol{\lambda}$, the differential of $\boldsymbol{\mathcal{L}}$ can be obtained from Eq.~\eqref{eq:lagrangian} with use made of \eqref{eq:chem_problem} to yield $\D\boldsymbol{\mathcal{L}}=0$, namely
\begin{equation}
    \rev{\left[\sum\limits_{j\in \mathbf{S}} \left(\mu_{j}(T, p, \mathbf{n}) + \sum\limits_{i\in \mathbf{E}} a_{ij} \lambda_i\right) \ +\ \sum\limits_{j\in \mathbf{S^C}} \varphi_j \right] {\rm d} n_j  + \sum\limits_{i\in \mathbf{E}}\left(\sum\limits_{j\in \mathbf{S}} a_{ij} n_j - b_i^\circ\right) {\rm d}\lambda_i = 0 },
    \label{eq:dlagrangian}
\end{equation}
\rev{which has to satisfy the complementary condition given in \eqref{eq:chem_problem_condensed}.} Additionally, the sum of the molar compositions must equal the total number of moles of gaseous species in the system, thus
\begin{equation}
    \label{eq:conservation_moles}
    \sum\limits_{j\in \mathbf{S^G}}n_j = n.
\end{equation}

Considering that ${\rm d}n_j$, ${\rm d}\lambda_i$, $\varphi_j$, and $n$ are independent, Eqs.~\eqref{eq:chem_problem_condensed}, \eqref{eq:dlagrangian}, and \eqref{eq:conservation_moles} constitute a system of NS non-linear equations subject to a set of NE + \rev{NC +} 1 linear constraints. Furthermore, to ensure that the molar composition $n_j$ is strictly positive, it is convenient to work with the functions $\ln n_j$ and $\ln n$ in the gaseous species. \rev{In our implementation, we found that excluding this approach for the condensed species resulted in a more efficient algorithm.}

Using these definitions, Eqs.~\eqref{eq:chem_problem_condensed}, \eqref{eq:dlagrangian}, and \eqref{eq:conservation_moles} can be rearranged in the form $\mathbf{f}(\mathbf{x}) = 0$, where $\mathbf{x}$ is the vector of unknowns composed of $\ln n_j^\text{\rev{G}}$, $n_j^\text{\rev{C}}$ \rev{and $\varphi_j^\text{\rev{C}}$}, $\ln n$, and $\pi_i = -\lambda_i/RT$. \rev{For the sake of conciseness, variables specific to gaseous species are denoted with the superscript G, indicating $\forall i \in \mathbf{S^G}$, while those related to condensed species are denoted with the superscript C, indicating $\forall j \in \mathbf{S^C}$.} To solve this system of equations, Combustion Toolbox uses a multidimensional Newton-Raphson (NR) method
\begin{equation}
    \mathbf{J} \cdot \delta\mathbf{x} = -\mathbf{f}(\mathbf{x}),
    \label{eq:NR_jacobian}%
\end{equation}%
where $\mathbf{J}$ is the Jacobian matrix with components $J_{ij} \equiv \partial f_i / \partial x_j$, $\delta\mathbf{x}$ is the correction vector composed of $\Delta\ln{n_j^\text{\rev{G}}}$, $\Delta n_j^\text{\rev{C}}$\rev{, $\Delta\varphi_j^\text{\rev{C}}$}, $\Delta\ln{n}$ and $\Delta\pi_i$, and $\mathbf{f}$ is the vector function. Equation \eqref{eq:NR_jacobian} can be expanded as the following system of NS \rev{+ NC} + NE + 1 linear equations
\begin{subequations}
    \begin{align}
        \Delta \ln n_j - \Delta \ln n - \sum\limits_{i \in\mathbf{E}} a_{ij}\Delta \pi_i &= -\dfrac{\mu_j}{RT}, \quad \forall j \in \rm \mathbf{S^G} \label{eq:EQUIL_Delta_nj}\\
        \sum\limits_{i \in\mathbf{E}} a_{ij}\Delta \pi_i \rev{\ +\ \Delta \varphi_j}  &= \dfrac{\mu_j}{RT} \rev{\ -\ \varphi_j} , \quad \forall j \in \rm \mathbf{S^C}\label{eq:EQUIL_Delta_nj_2}\\
        \rev{\varphi_j \Delta n_j + n_j \Delta \varphi_j}  &\rev{\ = \tau - n_j\varphi_j, \quad \forall j \in \rm \mathbf{S^C}}\label{eq:EQUIL_Delta_nj_5} \vphantom{\sum\limits_{}} \\
        \sum\limits_{j\in\mathbf{S^G}} a_{ij}n_j\Delta \ln{n_j} +\sum\limits_{j\in\mathbf{S^C}}a_{ij}\Delta n_j &= b_i^\circ - \new{\sum\limits_{j\in\mathbf{S}}} a_{ij}n_j,\quad \forall i \in \mathbf{E}\label{eq:EQUIL_Delta_nj_3}\\
       \sum\limits_{j\in\mathbf{S^G}} n_j\Delta\ln{n_j} - n \Delta\ln{n} &= n -\sum\limits_{j\in\mathbf{S^G}}n_j,\label{eq:EQUIL_Delta_nj_4}
    \end{align}%
    \label{eq:EQUIL_0}%
\end{subequations}%
where the dimensionless Lagrange multiplier $\pi_i$ has been taken equal to zero in the right hand side of Eqs.~\eqref{eq:EQUIL_Delta_nj} and \eqref{eq:EQUIL_Delta_nj_2}. This is a suitable simplification as long as $\lambda_i$ appears linearly in the first bracket of Eq.~\eqref{eq:dlagrangian}, as discussed in Ref.~\cite{zeleznik1968}. Algebraic manipulation of \eqref{eq:EQUIL_0} allows to reduce the system's dimensions due to the spareness of the upper left corner of the Jacobian matrix $\mathbf{J}$. Thus, substituting $\Delta \ln{n_j}$ from \eqref{eq:EQUIL_Delta_nj} in \eqref{eq:EQUIL_Delta_nj_3} and \eqref{eq:EQUIL_Delta_nj_4}\rev{, and $\Delta \varphi_j$ from \eqref{eq:EQUIL_Delta_nj_5} in \eqref{eq:EQUIL_Delta_nj_2}}, $\rm \rev{NS}$ equations are drop out from \eqref{eq:EQUIL_0}, providing a reduced \new{symmetric} system of $\rm NE\new{\ +\ NC}+1$ equations, namely
\begin{subequations}
    \begin{align}
        &\sum\limits_{i\in\mathbf{E}}\sum\limits_{j\in\mathbf{S}}{a_{lj}a_{ij}n_j\Delta \pi_i} + \new{\sum\limits_{j\in\mathbf{S^C}}{a_{lj}\Delta n_j}+\left(\sum\limits_{j\in\mathbf{S^G}}{a_{lj}n_j}\right)\Delta \ln{n}} = b_l^\circ-\sum\limits_{j\in\mathbf{S}} a_{lj}n_j+\sum\limits_{j\in\mathbf{S^G}}\frac{a_{lj}n_j\mu_j}{RT},\quad \forall l \in \mathbf{E} \label{eq:EQUIL_system_a}\\
        & \sum\limits_{i\in\mathbf{E}}{a_{ij} \Delta \pi_i} \rev{\ -\ \dfrac{\varphi_j}{n_j} \Delta n_j}=\frac{\mu_j}{RT} \rev{\ -\ \dfrac{\tau}{n_j}},\quad \forall j \in \mathbf{S^C} \label{eq:EQUIL_system_b}\\
        &\sum\limits_{i\in\mathbf{E}}\sum\limits_{j\in\mathbf{\new{S^G}}}{a_{ij}n_j\Delta \pi_i}+\left(\sum\limits_{j\in\mathbf{S^G}}{n_j-n}\right)\Delta\ln{n} 
        =n-\sum\limits_{j\in\mathbf{S^G}}n_j+\sum\limits_{j\in\mathbf{S^G}}\frac{n_j\mu_j}{RT}, \label{eq:EQUIL_system_c}
    \end{align}%
    \label{eq:EQUIL_system}%
\end{subequations}%
which is solved using MATLAB's \new{LU factorization with partial pivoting routine} built on LAPACK~\cite{anderson1999}. The updated solution vector $\mathbf{x}$ at the $(k+1)$-th iteration is given by  $\mathbf{x}_{k+1} = \mathbf{x}_{k} + \tau_k \, \delta \mathbf{x}_k$, where $\tau_k$ is the step size, or relaxation, parameter for the $k$-th iteration, defined as in \rev{Refs.~\cite{gordon1994,leal2016} for the gaseous and condensed phase, respectively}. As previously indicated, the solution vector $\mathbf{x}_{k}$ has $\pi_i = 0$; consequently, it is not necessary to relax the new value obtained, i.e., $\pi_{i, k+1} = \Delta\pi_{i, k}$.
Note that in Eq.~\eqref{eq:EQUIL_system}, $\mathbf{x}$ is composed of $n_j^\text{\rev{C}}$, $\ln n$, and $\pi_i = 0$. \new{Hence}, to update the terms $\ln{n_j^\text{\rev{G}}}$ \rev{and $\varphi_j^\text{C}$}, the corrections \rev{$\Delta \ln n_j^\text{G}$ and $\Delta \varphi_j^\text{C}$ must be obtained from \eqref{eq:EQUIL_Delta_nj} and \eqref{eq:EQUIL_Delta_nj_5}} after each solution of \eqref{eq:EQUIL_system}, which requires defining a set of initial estimates $\mathbf{x}_{0}$ for the molar number $n_j$ of all the possible products \rev{and for the slack variables $\varphi_j^\text{C}$}. \rev{We proceed using the max-min method~\cite{pope2003, pope2004} implemented in Mutation$^{++}$~\cite{scoggins2015,scoggins2020}. The initial composition is obtained from $\mathbf{n}_0 = (1 - \alpha) \mathbf{n}_0^{\rm major} + \alpha \mathbf{n}_0^{\rm minor}$ with a merging factor $\alpha = 0.01$. Here, $\mathbf{n}_0^{\rm major}$ represents the composition of the major chemical species and is obtained by solving Eq.~(\ref{eq:chem_problem}) and $\mathbf{n} \geq 0$ while considering $\mu \approx \mu^\circ$. The initial composition of the minor species $\mathbf{n}_0^{\rm minor}$ is defined as the composition that maximizes the smallest number of moles of a single species while still satisfying (\ref{eq:chem_problem}). Both problems are solved using simplex algorithms (see \texttt{simplex.m} and \texttt{simplexDual.m} routines, respectively).} However, when performing parametric sweeps, CT starts with the moles of the gaseous species $n_j$ obtained from the previous calculation, provided that their magnitudes exceed a predefined threshold value of $10^{-6}$. This approach facilitates the convergence of the iterative procedure and accelerates the overall computational efficiency.

\rev{The algorithm first solves the ECP problem for the gas phase, removing species with molar compositions below a certain tolerance to prevent an ill-conditioned $\mathbf{J}$ matrix.} Once convergence is achieved (by default, the tolerance for the molar composition is set to $10^{-14}$ and for the NR is $10^{-5}$), if there are condensed species in the set of products, a second iteration process is conducted. The procedure is similar, but now we include in the set of unknowns \rev{all} the condensed species that satisfy the vapor pressure test, namely
\begin{equation}
     \frac{1}{RT} \frac{\partial \mathcal{L}}{\partial n_j} = \frac{\mu_j^\circ}{RT}-\sum\limits_{i\in\mathbf{E}}{\Delta \pi_ia_{ij}} < 0, \quad \forall j \in \mathbf{S^C}
    \label{eq:condition_condensed}
\end{equation}
whose addition to the system will reduce the Gibbs free energy of the system even further, corresponding with the first bracket of Eq.~\eqref{eq:dlagrangian} in dimensionless form\rev{. The term $\varphi_j^{\text{C}}$ is omitted because it is effectively zero when the condensed species is stable}. Note that if \eqref{eq:condition_condensed} yields negative values, the Lagrange function may not be at equilibrium ($\D \boldsymbol{\mathcal{L}} = 0$), which means that the added species can appear at the final state of equilibrium. \rev{If, in the new equilibrium state, the slack variable yields $|\log[\exp{(-\varphi_j^\text{C} / RT)}]| > 10^{-2}$, the molar composition $n_j^\text{C}$ of the added species drops below the tolerance threshold, or the Jacobian matrix $\mathbf{J}$ becomes singular, these species are considered unstable at equilibrium and are consequently removed from the set~$\mathbf{S^C}$.} \new{This} process is repeated until all the condensed species in $\mathbf{S^C}$ that satisfy $T\in \left[T_{\rm min}, T_{\rm max}\right]$, i.e., whose temperature range is compatible with the system's temperature, have been tested. \rev{For any added condensed species whose initial compositions were not predefined from the max-min method, we assign a starting value of $\mathbf{n_0^C} = 10^{-6}$. Similarly, for the slack variables, we initially assume that all species are stable, setting their values to $\mathbf{\varphi_0^C} = 10^{-14}$.}

In the presence of ionized gases, the algorithm neglects the Coulombic interactions associated with ideal plasmas. In this case, there is only an additional restriction that is given by the electroneutrality of the mixture~\cite{smirnov2012}
\begin{equation}\label{electro_neutrality}
    \sum\limits_{j\in\mathbf{S^G}} a_{ej} n_{j} = 0,
\end{equation}
where the stoichiometric coefficient $a_{ej}$ represents the number of electrons in ion $j$ relative to the neutral species. Thus, for $\{\rm{e}^-, \rm{N}_2^+\}$ we have $a_{ej} = \{1, -1\}$. This means that the electron $E_e$, with index $e$, is treated as an element \rev{and has an associated Lagrange multiplier $\pi_e$ ($\lambda_e$) in dimensionless form (dimensional form)}. This assumption is valid only when the ion density is sufficiently small, i.e., for weakly ionized gases. The code directly detects if there are ions in the set of possible products $\mathbf{S}$ and calls another subroutine that ensures that condition \eqref{electro_neutrality} is met.

\subsection{Equilibrium composition at specified temperature and volume (TV)}
\label{sec:EQUIL_kernel_2}

For calculating the molar equilibrium composition of the mixture at a given temperature $T$ and volume $v$, we have to minimize the Helmholtz free energy of the system, defined as $F = G - pv$. Upon use of this relation into the minimization condition ${\rm d}F\left(T,v,\mathbf{n}\right) = 0$, we get
\begin{equation}
    \sum\limits_{j\in\mathbf{S}} \mu_j(T, v, \mathbf{n})\; n_j - pv = 0
\end{equation}
to be used in substitution of Eq.~\eqref{eq:min_gibbs_a}. For convenience, the chemical potential of species $j$ is now expressed as a function of the mixture's volume $v$. For an ideal gas EoS we have
\begin{equation}
\mu_j\left(T,v,\mathbf{n}\right)=\mu_j^\circ\left(T\right)+\kappa_j RT\left(\ln{\frac{n_j}{\sum\limits_{j\in\mathbf{S^G}} n_j}}+\ln{\frac{n_j R T}{p^\circ v}}\right), \quad \forall j \in \mathbf{S}
    \label{eq:chemical_potential_TV}
\end{equation}
which ultimately gives a different reduced system of equations\new{,} to be solved instead of \eqref{eq:EQUIL_system}, namely
\begin{subequations}
    \begin{align}
        &\sum\limits_{i\in\mathbf{E}}\sum\limits_{j\in\mathbf{S}}{a_{lj}a_{ij}n_j\Delta \pi_i}+\sum\limits_{j\in\mathbf{S^C}}{a_{lj}\Delta n_j} = b_l^\circ-\sum\limits_{j\in\mathbf{S}} a_{lj}n_j+\sum\limits_{j\in\mathbf{S^G}}\frac{a_{lj}n_j\mu_j}{RT},\quad \forall l \in \mathbf{E}\\
        & \sum\limits_{i\in\mathbf{E}}{a_{ij} \Delta \pi_i} \rev{\ -\ \dfrac{\varphi_j}{n_j} \Delta n_j}=\frac{\mu_j}{RT} \rev{\ -\ \dfrac{\tau}{n_j}},\quad \forall j \in \mathbf{S^C}.
    \end{align}
    \label{eq:EQUIL_system2}%
\end{subequations}
Unlike in the TP calculations, the linear system no longer includes the total number of moles, $n$, and its correction factor, $\Delta \ln{n}$, as they are drop out of the system of $\rm NE \new{\ +\ NC}$ dimensions written above. Here, $\mu_j$ is given by Eq.~\eqref{eq:chemical_potential_TV} and the correction values $\Delta \ln n_j^\text{\rev{G}}$ do not depend of $\Delta \ln{n}$, yielding
\begin{equation}
    \Delta \ln n_j = \sum\limits_{i \in\mathbf{E}} a_{ij}\Delta \pi_i -\dfrac{\mu_j}{RT}, \quad \forall j \in \rm \mathbf{S^G}.
    \label{eq:EQUIL_Delta_nj_TV}
\end{equation}
\rev{On the other hand, the values of $\Delta \varphi_j^\text{C}$ are provided by the same expression \eqref{eq:EQUIL_Delta_nj_5}.} The computation of the chemical composition and thermodynamic properties of a given mixture at specified temperature and volume is performed by the routine \rev{\texttt{equilibriumHelmholtz}}.

\subsection{Equilibrium composition for other pairs of state functions}
\label{sec:EQUIL_remainder}

In many practical applications, the equilibrium temperature of \new{the system is not known \textit{a-priori}, which requires} the provision of supplementary information to close the problem. This additional information may be obtained from an enthalpy, internal energy, or entropy conservation equation, subject to the requirement that the corresponding state function $f$ remains unchanged, namely
\begin{equation}
     \Delta f\left(T\right) \equiv f_{\rm F}\left(T\right) - f_{\rm I}\left(T_{\rm I}\right) = 0,
     \label{eq:root}
\end{equation}
where the subscripts F and I refer here to the final and initial states of the mixture, respectively. Unlike in NASA's CEA code, we have increased the flexibility of the CT-EQUIL module by decoupling this additional equation and retrieved the new condition by using a second-order NR method
\begin{equation}
    T_{k+1} = T_k-\frac{f\left(T_k\right)}{f^\prime\left(T_k\right)}.
    \label{eq:NR}
\end{equation}

The derivatives of the state functions $f^\prime\left(T\right)$ involved in the different transformations can be expressed analytically in the form: $\left(\partial h/\partial T\right)_p=~c_p$, $\left(\partial e/\partial T\right)_v=~c_v$, $\left(\partial s/\partial T\right)_p=~c_p/T$, and $\left(\partial s/\partial T\right)_v=~c_v/T$, for HP, EV, SP, and SV transformations, respectively. Following common practice, $h$, $e$, $s$, $c_p$, and $c_v$ denote \new{here} the enthalpy, internal energy, entropy, and the specific heats at constant pressure and constant volume, respectively. It is worth noting that that although they are written in lower case letters, these variables refer here to extensive magnitudes.

The initial estimate $T_0$ is computed using a \textit{regula falsi} method. Nevertheless, when carrying out parametric studies, the program uses the temperature obtained in the previous calculation as an initial estimate to accelerate convergence toward the new solution. However, if $T_k$ is significantly distant from the actual solution, this approach can lead to unsatisfactory convergence. To overcome this issue, we can select a more robust root-finding method, as Eq.~\eqref{eq:root} has been purposely decoupled. In particular, the code has implemented an implicit third-order Newton-Steffensen root-finding algorithm~\cite{sharma2005}, defined as follows:
\begin{equation}
T_{k+1} = T_k-\frac{f^2\left(T_k\right)}{f^\prime\left(T_k\right)\left[f\left(T_{k}\right) - f\left(T^*_{k+1}\right)\right]},
\label{eq:steff}
\end{equation}
where the temperature at the $(k+1)$-th iteration, $T_{k+1}$, is re-estimated by using the provisional value $T^*_{k+1}$ provided by the application of the classical method defined in \eqref{eq:NR}. The convergence criterion $\max\{|(T_{k+1} - T_{k})/T_{k+1}|, |\Delta f / f_F|\} < \epsilon_{0}$ is set by default to $10^{-3}$ in both methods and is generally reached in two to five iterations.

\subsection{Validations}
\label{sec:equil_validations}
To illustrate the wide variety of applications of Combustion Toolbox and asses the capabilities of the CT-EQUIL module, several validation tests have been conducted. In this work, we provide three of them in which only a reduced set of species are presented for clarity. Further validation tests can be easily accessed through the CT website or by just utilizing the user-interface validation add-on, \textit{uivalidation}, which is implemented in the GUI. Alternatively, the user can also run the scripts included in the \textit{validations} folder.

\begin{figure}[!htpb]
    \centering
    \includegraphics[width=0.94\textwidth]{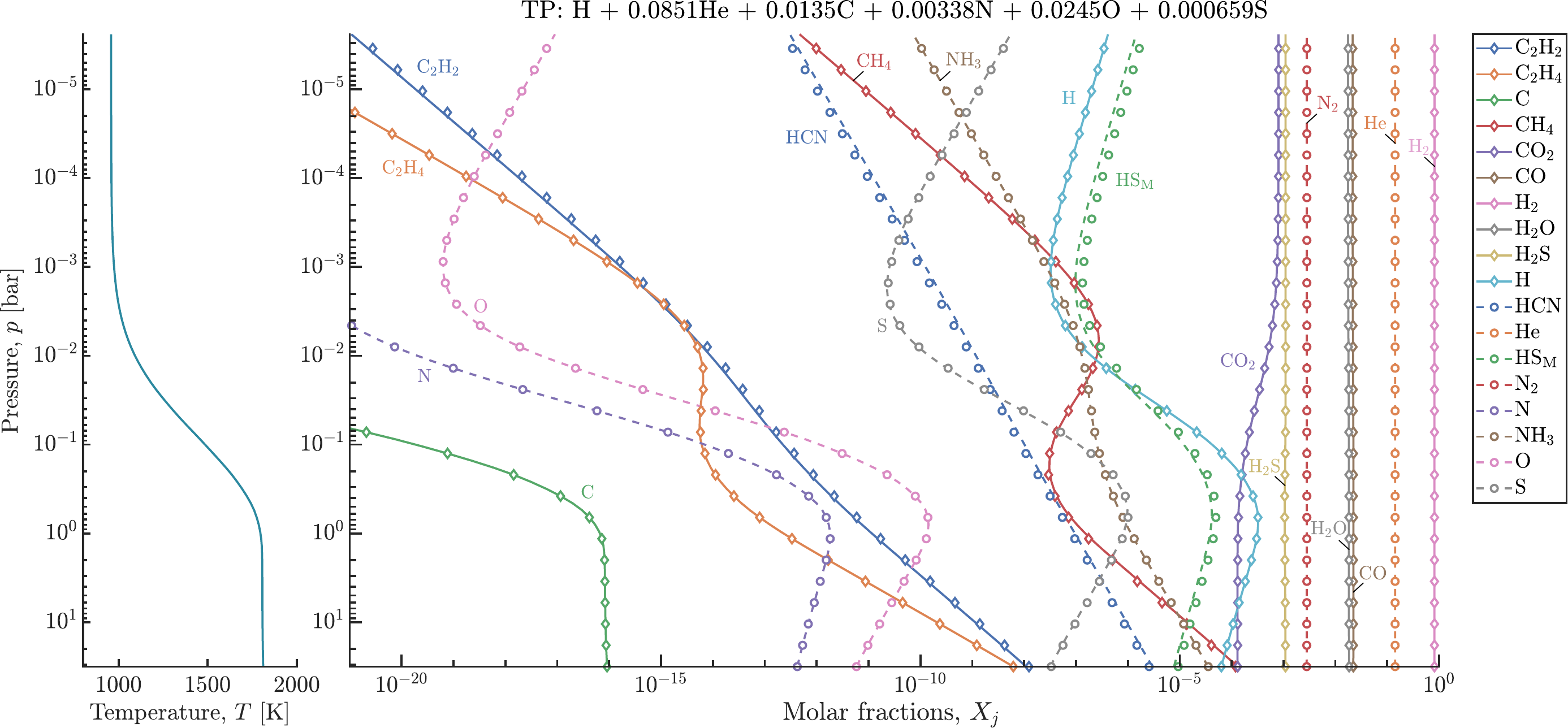}
    \caption{Variation of molar fraction with pressure (right panel) for the temperature-pressure profile of exoplanet WASP-43b (left panel) with an atmospheric $50\times$ solar metallicity; solid line: numerical results obtained with CT; symbols: numerical results obtained with TEA~\cite{blecic2016}. The species denoted with subscript M is obtained from Burcat's database~\cite{burcat2005}.}
    \label{fig:results_exoplanets}
\end{figure}

\textit{First test:} In planetary science, thermochemical equilibrium codes like TEA~\cite{blecic2016}, Fastchem~\cite{stock2018, stock2022, kitzmann2024}, and GG$_{\rm CHEM}$~\cite{woitke2018}, are used to model the atmospheric composition of giant planets, brown dwarfs, and other celestial bodies. Further examples can be found in Refs.~\cite{fegley1996, visscher2006, agundez2014a, parmentier2018}. Such models can help to \new{reveal} the physicochemical processes (\new{both} chemical and radiative) that drive the evolution of these atmospheres, \new{including} the formation of clouds and the escape of atmospheric gases into space~\cite{madhusudhan2019}. This motivates the first validation case: the composition of the hot-Jupiter exoplanet WASP-43b's atmosphere. To this end, it is necessary to provide temperature and pressure profiles, as well as the planet's metallicity, which refers to the abundance of elements heavier than hydrogen and helium present in its composition. For example, Fig.~\ref{fig:results_exoplanets} shows the variation of the species molar fractions $X_j=n_j/\sum_{j\in\mathbf{S}} n_j$ with pressure (right panel) corresponding to the temperature-pressure profile given in~\cite{stevenson2014} (left panel), and assuming an atmospheric $50\times$ solar metallicity. This value is needed to determine the mixture's initial number of moles $n_j$, upon knowledge of the elemental solar abundances that are here estimated from \cite{asplund2009}, which considers H the reference element. \rev{The \texttt{readAbundances} method, encapsulated within the \texttt{SolarAbundances} class, facilitates the retrieval of solar mass abundances. These quantities are then processed into molar abundances through the class method \texttt{abundances2moles}, where a default unity metallicity is assumed.}

It is important to note that different exoplanets may necessitate the utilization of distinct solar abundances. Nevertheless, the program is compatible with additional datasets that follow the same format as the original dataset \texttt{abundances.txt} (located in the \textit{databases} folder). The results (solid lines) show an excellent agreement with those obtained with the Thermochemical Equilibrium Abundances (TEA) code~\cite{blecic2016} (symbols) even down to the $\mu$bar level. The minor differences in HCN, C$_2$H$_2$ (acetylene), and HS$_{\rm M}$ (the subscript M denotes that is obtained from Burcat's database) come from the discrepancies of the free energies compared to the NIST-JANAF database~\cite{chase1998, dorofeeva2001} that is implemented in TEA. In this test, the computation time with a tolerance of $10^{-32}$ for the molar composition was \rev{0.9} seconds (TEA: 6.42 seconds) for a set of 26 species considered and a total of 90 case studies, which represents a \rev{$7\times$} speed-up factor for our code (\rev{$36\times$} loading a specific database for this case).

\begin{figure}
    \centering
    \includegraphics[width=0.94\textwidth]{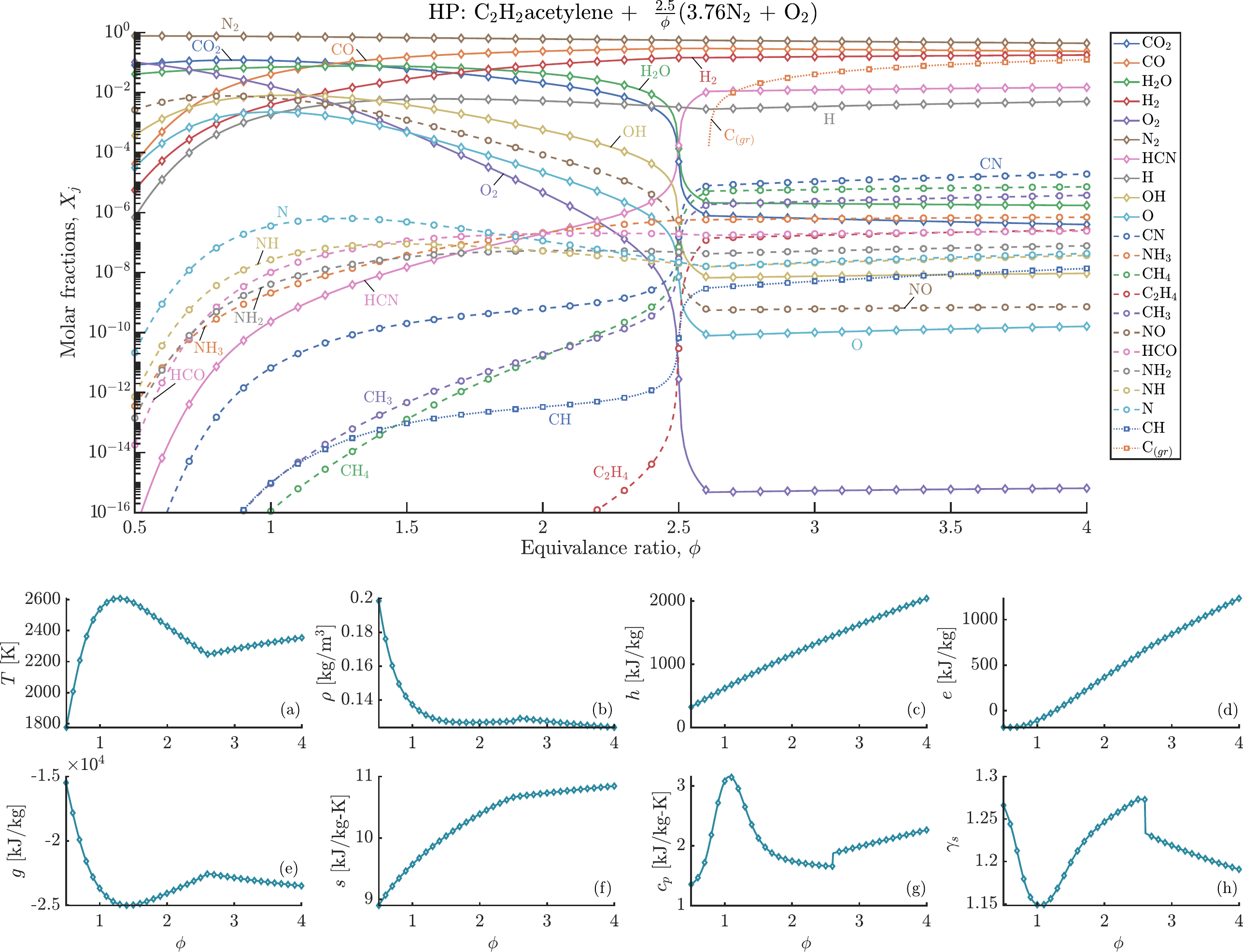}
    \caption{Variation of the molar fractions, $X_j$ (top), and of different thermodynamic mixture properties: (a) temperature, $T$, (b) density, $\rho$, (c) enthalpy, $h$, (d) internal energy, $e$, (e) Gibbs energy, $g$, (f) entropy, $s$, (g) specific heat capacity at constant pressure, $c_p$, and (h) adiabatic index, $\gamma_s$, for an HP transformation in lean-to-rich acetylene (C$_2$H$_2$)-air mixtures at standard conditions ($T_1 = 300$ K, $p_1 = 1$ bar); solid line: numerical results obtained with CT; symbols: numerical results obtained with NASA's CEA \new{code}~\cite{gordon1994}.}
    \label{fig:results_HP}
\end{figure}
\begin{figure}[!htpb]
    \centering
    \includegraphics[width=0.93\textwidth]{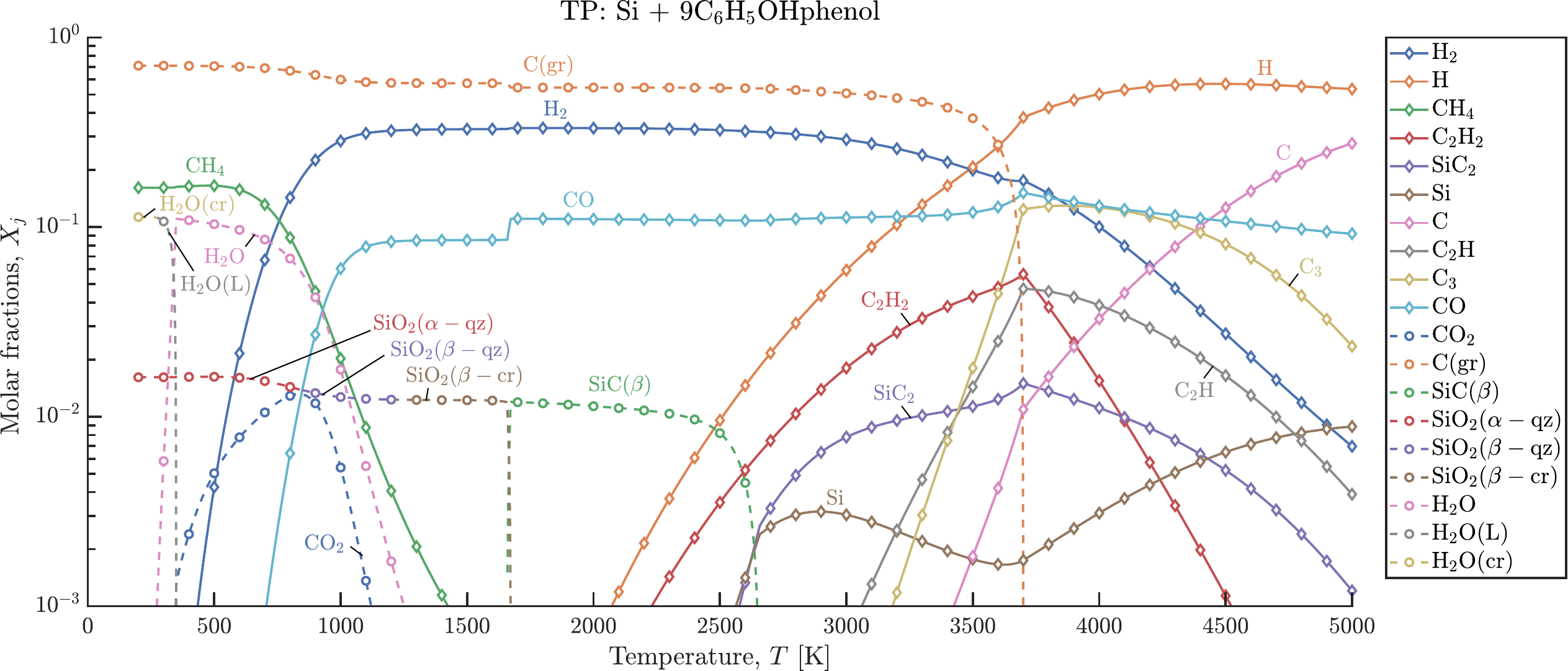}
    \caption{Variation of the molar fractions $X_j$ for a Silica-Phenolic mixture at atmospheric pressure ($p = 1$ atm) with $T \in [200, 5000]$; solid line: numerical results obtained with CT; symbols: numerical results obtained with NASA's CEA \new{code}~\cite{gordon1994}.}
    \label{fig:results_TP_scoggins2015}
\end{figure}

\textit{Second test:} As previously indicated, thermochemical codes are \new{essential for understanding and predicting the intricate chemical transformations that take place in combustion processes. Therefore, validating CT with a canonical combustion test is critically important. To this end, the second validation test involves the adiabatic isobaric combustion of acetylene and air, a promising mixture for advanced internal combustion engines due to acetylene's} high energy density and low carbon content. Specifically, the investigation focuses on the isobaric reactive mixture at an initial pressure of $p_1 = 1$ atm and temperature of $T_1 = 300$ K, and considers a wide range of equivalence ratios $\phi \in [0.5, 4]$. Figure~\ref{fig:results_HP} shows the variation of the molar composition of the products with $\phi$ (top panel) along with other mixture properties (Figs.~\ref{fig:results_HP}a-h).
It should be noted that, unlike the previous test, in this case the results obtained with CT (represented by solid lines) are compared to those of NASA's CEA \new{code}~\cite{gordon1994} (represented by symbols). Once again, the results are in excellent agreement with the benchmark code, \new{including} the generation of solid carbon C$_{\rm(gr)}$ when the equivalence ratio reaches or exceeds $\phi\approx2.6$. The computation time elapsed \rev{3.34} seconds for a set of 94 species and 351 case studies\rev{, which is $1.4\times$ faster compared to the previous release~\cite{combustiontoolboxv1, cuadra2023_thesis}.}

\textit{Third test:} Recent advancements in chemical equilibrium solvers have opened up new opportunities for studying the complex phenomena that take place on the surface of ablative materials during atmospheric reentry. For instance, Helber et al.~\cite{helber2014} used a chemical equilibrium code to investigate the ablation of carbon-based materials under conditions experienced during Earth's atmospheric reentry. Other studies focused on carbon-fiber-reinforced-polymers (CFRP)~\cite{bariselli2020} or silicon-based ablative materials~\cite{park2021}. Regardless of the composition of the ablator, predicting the equilibrium species \new{generated during the ablation process} is a challenging task, as it involves the evaluation of \new{a large number of} condensed species at very high enthalpies. To assess the capabilities of CT under more demanding scenarios than those of the previous tests, we recall in this third test the example presented in Ref.~\cite{scoggins2015}. The problem consists of a parametric study of a Si-C$_6$H$_5$OH mixture at atmospheric pressure ($p = 1$ atm) for a wide range of temperatures $T \in [200, 5000]$ \new{K}. \new{Figure~\ref{fig:results_TP_scoggins2015} illustrates the variation with temperature of the equilibrium molar fractions $X_j$ for the Silica-Phenolic mixture, along with the results computed using NASA's CEA code~\cite{gordon1994}}. It \new{can be} seen that there is total agreement even for the multiple condensed species. Previous work~\cite{scoggins2015} reported that NASA's CEA code was not able to converge for $T < 400$ \new{K}. However, this is only true when computing the parametric study, not the individual cases, whereas CT converges in both situations. In this test, the computation time was \rev{3.05 seconds for a set of 177 species (40 in condensed phase)} and 481 case studies\rev{, making this version 1.67$\times$ faster than the previous release~\cite{combustiontoolboxv1, cuadra2023_thesis}.}

\section{Shock and detonation module}
\label{sec:SD}
This section presents the routines of the shock and detonation module, CT-SD\rev{, which are encapsulated in the \texttt{ShockSolver} and \texttt{DetonationSolver} classes, respectively. These solvers depend on the CT-EQUIL module described above, and therefore, each object is associated with an instance of the \texttt{EquilibriumSolver} class}. \new{The} CT-SD module determines the post-shock equilibrium state of steady non-reactive and reactive shocks with arbitrary incidence angles, $\beta$ (see inset on the right panel of Fig.~\ref{fig:shock_polar}b). The \new{methods to solve normal shocks and detonation waves, $\beta = \pi/2$,} are based on the algorithm outlined in NASA's Reference Publication 1311~\cite[Chapters 7-9]{gordon1994}.

\subsection{Oblique shocks}
\label{sec:oblique_shocks}

\new{First,} consider the problem of an undisturbed, planar, normal shock wave. The pre-shock density, pressure, enthalpy, and velocity (in the reference frame attached to the shock) are denoted, respectively, as $\rho_1$, $p_1$, $h_1$, and $u_1$. The corresponding flow variables in the post-shock gases are denoted as $\rho_2$, $p_2$, $h_2$, and $u_2$. The well-known Rankine-Hugoniot (RH) relations for the variation of pressure and enthalpy are, respectively
\begin{equation}
    p_2 = p_1 + \rho_1 u_1^2 \left( 1-\dfrac{\rho_1}{\rho_2}\right) \quad \new{\text{and}} \quad 
    h_2 = h_1 + \dfrac{u_1^2}{2}\left[1- \left(\dfrac{\rho_1}{\rho_2}\right)^2\right].
\label{eq:RH}
\end{equation}
These equations must be supplemented by the equation of state, in our case, the ideal EoS $p=\rho R T/W$, where $W = \sum_{j\in\mathbf{S^G}} (n_j /\sum_{j\in\mathbf{S^G}} n_j) W_j$ stands for the average molecular mass of the gaseous mixture computed in terms of the gas-phase molar fractions, $n_j/\sum_{j\in\mathbf{S^G}} n_j$, and the molecular masses of the gaseous species, $W_j$. Also required is the caloric EoS, $h = \sum_{j\in\mathbf{S}} n_j H_{j}^{\circ}(T)$, that gives enthalpy in terms of the temperature and gas mixture composition. As \new{previously} discussed, the molecular masses, $W_j$, and the molar specific enthalpies, $H_{j}^{\circ}(T)$, are evaluated from a combination of NASA's~\cite{Mcbride2002} and Burcat's (Third Millennium)~\cite{burcat2005} thermochemical databases. 

\begin{figure}[!ht]
    \centering
    \includegraphics[width=0.85\textwidth]{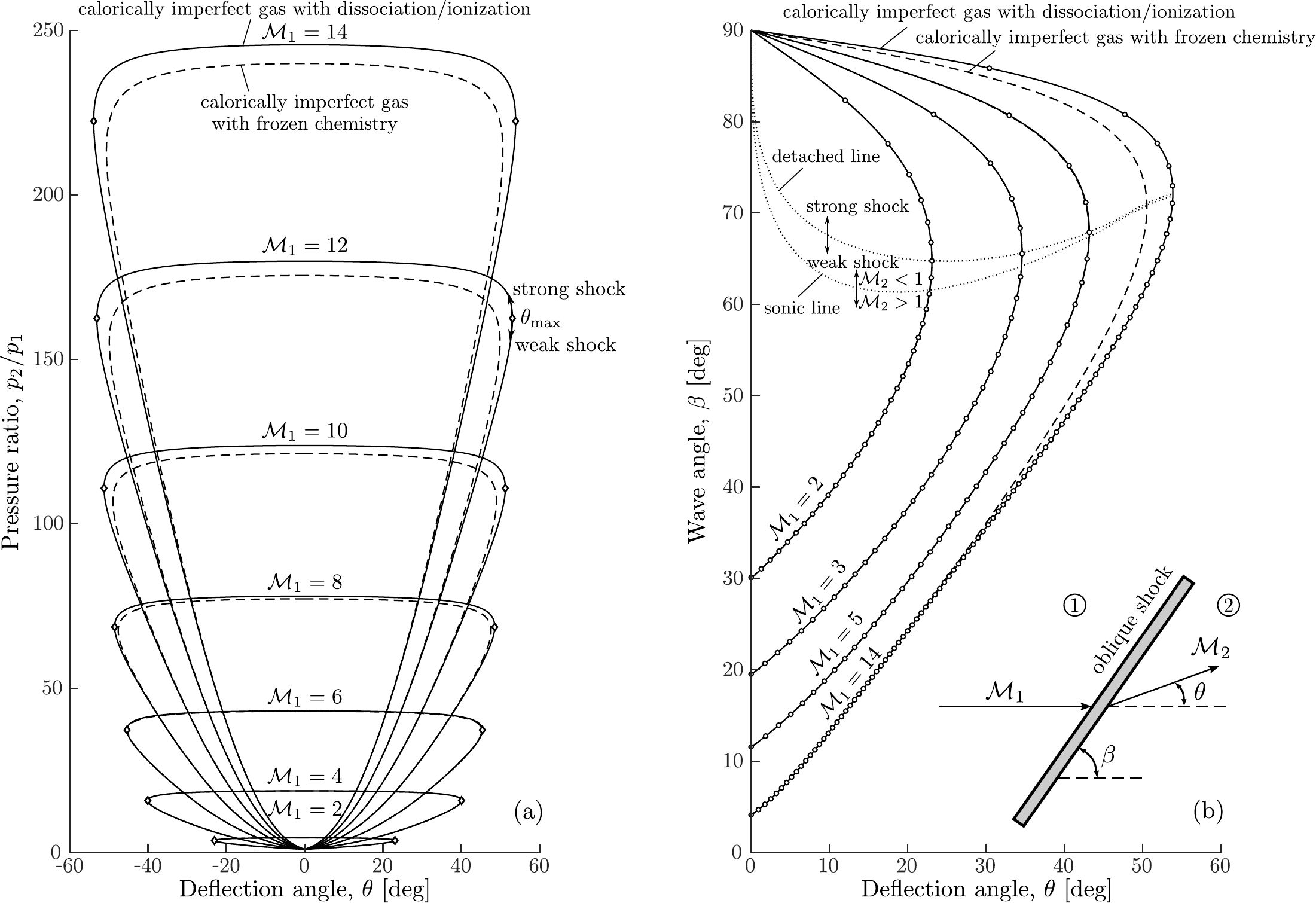}
    \caption{Pressure-deflection (a) and wave angle-deflection (b) shock polar diagrams for air (78\% N$_2$, 21\% O$_2$, and 1\% Ar) at pre-shock temperature $T_1 = 300$ K and pressure $p_1 = 1$ atm, and a range of pre-shock Mach numbers $\mathcal{M}_1$ between 2 and 14; solid line: calorically imperfect gas with ionization/dissociation; dashed: calorically imperfect gas with frozen chemistry; circles: results obtained with Cantera~\cite{Cantera} within Caltech's SD-Toolbox~\cite{Browne2008}; diamonds: maximum deflection angle $\theta_{\rm max}$.}
    \label{fig:shock_polar}
\end{figure}

The solver employs a NR method (see Ref.~\cite[Chapters 7-9]{gordon1994} for more information) to determine the roots of the system of equations governing both reactive and non-reactive shocks (routines \rev{\texttt{detCJ}} and \rev{\texttt{shockIncident}}, respectively). Consequently, the jump relationships, such as $p_2/p_1$ and $\rho_2/\rho_1$, can be calculated with ease. In this subsection, we focus on the oblique shock configuration that implicitly includes the solution of the normal shock wave. When the angle formed by the shock plane and the upstream flow is not $\pi/2$, in the RH equations \eqref{eq:RH} the pre- and post-shock velocities, $u_1$ and $u_2$, must be replaced by their respective components normal to the shock, $u_{1n}$ and $u_{2n}$. These equations can be readily reformulated in terms of the magnitudes of the pre- and post-shock velocities upon direct substitution of the trigonometric relationships $u_{1n} = u_1 \sin{\beta}$ and $u_{2n} = u_2 \sin{\left(\beta - \theta\right)}$, where $\beta$ is the shock incidence angle and $\theta$ is the flow deflection angle, both measured with respect to the upstream flow direction. For oblique shocks, the continuity of the tangential velocity across the shock, $u_{1t} = u_{2t}$, or, equivalently,  $u_1 \cos{\beta}= u_2 \cos{\left(\beta - \theta\right)}$, is also required.

In the case of a normal shock wave, the gas properties downstream of the shock are determined by two factors: the upstream thermodynamic state and a parameter that characterizes the intensity of the shock, usually either the shock speed $u_1$ relative to the upstream gas or the shock Mach number, $\mathcal{M}_1=u_1/a_1$\new{,} where $a_1$ represents the speed of sound upstream the shock\new{. The speed of sound, $a$, is formally defined as $a^2 = p / \rho \left(\partial \ln p / \partial \ln \rho\right)_s $}. Other problems, such as those concerning blast waves, impose the post-shock pressure $p_2$ as the initial input parameter describing the shock intensity. Oblique shocks, on the other hand, require an additional geometrical restriction given by the value of the incidence angle $\beta$ or the flow deflection angle $\theta$. \new{It is well known that} when the value of $\beta$ is specified, there exists a unique solution for the post-shock fluid state (see routine \rev{\texttt{shockObliqueBeta}}). However, in the case where $\theta$ is given, there are two possible solutions for $\beta$: a weak shock solution linked to a smaller value of $\beta$, representing the weakest possible shock, and a strong solution corresponding to a larger $\beta$ (see routine \rev{\texttt{shockObliqueTheta}})\new{~\cite{Pratt1991}}. The two solutions converge for $\theta = \theta_{\rm max}$, which corresponds to the maximum deflection angle of the shock for a given ${\cal M}_1$ (see Fig.~\ref{fig:shock_polar}), above which the problem does not admit \new{any} solution. Remarkably close to the maximum deflection angle, in the weak-shock branch, we find the sonic condition $\mathcal{M}_2 = 1$, below and above which the post-shock flow is supersonic (weak shocks) and subsonic (strong shocks, and weak shocks between the sonic line and the maximum deflection angle), respectively. Then, the supersonic branch always corresponds to the weak shock solution\new{~\cite{shapiro1953}}. 

In either scenario, when the value of $\theta$ is specified, $\beta$ becomes an implicit variable that must be determined numerically. To this end, CT employs an iterative procedure based on the continuity of the tangential velocity across the shock, which can be manipulated using the above trigonometric identities to give
\begin{equation}
    f\left(\beta\right)\equiv\theta+{\tan}^{-1}{\left(\frac{u_{2n}}{u_1\cos{\beta}}\right)}-\beta = 0.
    \label{eq:oblique_root}
\end{equation}
This equation must be solved for the shock incidence angle $\beta$ with use made of the RH relations \eqref{eq:RH}, the ideal gas EoS, and provided that $f^\prime\left(\beta\right)$ and $f^{\prime\prime}\left(\beta\right)$ can be written as explicit functions. This enables the use of Halley's third-order iterative method~\cite{scavo1995}
\begin{equation}
    \beta_{k+1} = \beta_k - \frac{2 f\left(\beta_k\right) f^\prime\left(\beta_k\right)}{2 f^\prime\left(\beta_k\right) - f\left(\beta_k\right)f^{\prime\prime}\left(\beta_k\right)}
    \label{eq:Halley}%
\end{equation}
to find the root of Eq.~\eqref{eq:oblique_root}. This approach exhibits rapid convergence and meets the default convergence criterion of $10^{-3}$ within two or three iterations. However, it is worth noting that like other root-finding methods, Halley's method only provides one of the possible roots of the nonlinear system, which generally corresponds to the root closest to the initial guess used during the iterative process. Therefore, we must supply sufficiently accurate guesses to cover the whole set of solutions that include both the weak- and strong-shock branches. One straightforward approach to acquiring such guesses is to anticipate the solution domain bounded by the acoustic weak-shock limit $\beta_{\min}={\sin}^{-1}{\left(1/\mathcal{M}_1\right)}$ and the normal shock configuration $\beta_{\max}=\pi/2$. In particular, we choose $\beta_0=0.5\left(\beta_{\min}+\beta_{\max}\right)$ for the weak-shock branch and $\beta_0=0.97\beta_{\max}$ for the strong-shock branch.

The left and right plots in Fig.~\ref{fig:shock_polar} depict the pressure ratio-deflection angle and the incidence angle-deflection angle shock polar diagrams for dry air (consisting of 78\% N$_2$, 21\% O$_2$, and 1\% Ar) initially at room conditions ($T_1 = 300$ K, $p_1 = 1$ atm). These results were obtained using the \rev{\texttt{shockPolar} method}. It is worth noting that CT provides users with a variety of \new{caloric} models to choose from: \emph{i)} calorically perfect gas with frozen chemistry (constant specific heat at constant pressure $c_p$, adiabatic index $\gamma$, and $n_j$), \emph{ii)} calorically imperfect gas with frozen chemistry (constant $n_j$), and \emph{iii)} calorically imperfect gas with variable composition, including dissociation, ionization, and recombination reactions at equilibrium. These \new{models} are incorporated by specifying a sufficiently large set of species for the calculations, and using NASA's~\cite{Mcbride2002} and Burcat's (Third Millennium)~\cite{burcat2005} databases for evaluating the thermodynamic properties.

Figure~\ref{fig:shock_polar} displays the results obtained with models \emph{ii)} and \emph{iii)} spanning a set of pre-shock Mach numbers $\mathcal{M}_1$ ranging from 2 to 14. The results are compared with Caltech's Shock and Detonation Toolbox~\cite{Browne2008SDT}, which uses Cantera~\cite{Cantera} as kernel for the computations of chemical equilibrium. It is found that the lobes in the pressure ratio-deflection angle diagram expand due to dissociation/ionization effects, particularly in the hypersonic flow regime, $\mathcal{M}_1 > 5$. As a result, weak oblique shocks exhibit smaller pressure ratios while strong ones exhibit larger pressure ratios for the same deflection angle. Moreover, the endothermic (cooling) effect caused by dissociation/ionization in hypersonic oblique shocks leads to an increase in the post-shock density that also increases the wave deflection angle at all incidence angles. The results obtained from both codes are in complete agreement for all conditions tested. However, CT-SD exhibits superior performance compared to Caltech's SD-Toolbox with Cantera. \rev{Our code performs $26\times$ faster on average} (CT-SD: \rev{3.78} \new{seconds} vs. Caltech's SD-Toolbox \& Cantera: 99.72 \new{seconds}; for a large subset that contains 1200 points of all the cases represented in Fig.~\ref{fig:shock_polar}b, with both codes running on the same platform and with the same subset of \new{13} chemical species), which demonstrates the excellent performance of the CT-SD module.

\subsection{Regular reflections}
\label{sec:regular_reflections}

\begin{figure}[!ht]
    \centering
    \includegraphics[width=0.834\textwidth]{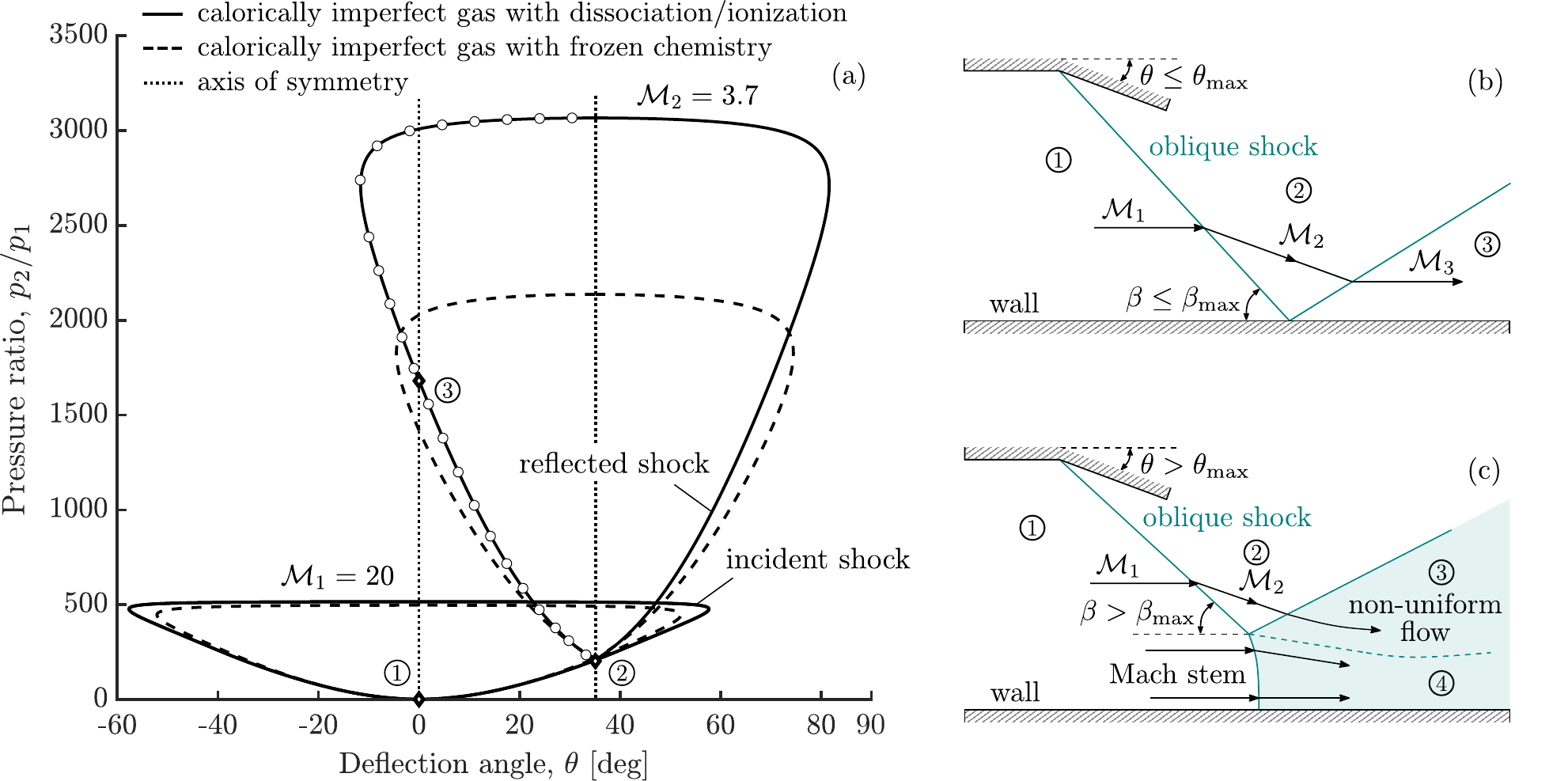}
    \caption{\new{Pressure-deflection shock polar diagrams for a regular shock reflection in atmospheric air (78\% N$_2$, 21\% O$_2$, and 1\% Ar) at 30 km above sea level (pre-shock temperature $T_1 = 226.51$ K and pressure $p_1 = 1.181\cdot10^{-2}$ atm), pre-shock Mach number $\mathcal{M}_1 = 20$, and deflection angle $\theta = 35^\circ$; solid line: calorically imperfect gas with dissociation/ionization; dashed line: calorically imperfect gas with frozen chemistry; dotted line: axes of symmetry; circles: results of Zhang et al.~\cite{zhang2022a}; diamonds: states 1, 2, and 3.}}
    \label{fig:validation_shocks_RR}
\end{figure}

Understanding the reflection of shock waves off flat surfaces is a problem of great relevance to high-speed flows. The angle subtended by the incident shock and the flat surface determines the type of shock reflection, with $\beta = 0$ representing normal reflections and $0<\beta<\pi/2$ oblique reflections. In a reference frame with origin at the point of contact of the shock with the wall, the latter exhibit an incoming free stream parallel to the wall, with $\mathcal{M}_1 > 1$. For sufficiently small incidence angles, the incident shock deflects the free stream uniformly towards the wall an angle $\theta$. The reflected shock then deflects back the perturbed stream to its original flow direction parallel to the wall. This type of reflection is \new{called} regular, and is depicted in Fig.~\ref{fig:validation_shocks_RR}\new{b}. Regular reflections leave uniform flow patterns behind both the incident and reflected shocks\new{, which can be thoroughly computed using} Combustion Toolbox.

By contrast, for incidence angles above a certain critical value, $\beta > \beta_{\max} (\mathcal{M}_1)$, the reflected wave, \new{characterized by} $\mathcal{M}_2 < \mathcal{M}_1$ and \new{consequently} a lower maximum deflection angle, is not able to deflect the flow back to its upstream direction parallel to the wall. This leads to the so-called irregular, or Mach, reflections, where the reflected and incident shocks merge into a single wave called the Mach stem, which connects the wall to the triple point where the three shocks meet\new{, as illustrated in Fig.~\ref{fig:validation_shocks_RR}c}. These reflections produce non-uniform flows that include a high-speed shear layer, or slipstream, emanating from the triple point~\cite{Hornung1986}. The properties of these flows cannot be determined solely by the polar-plot charts or the zero-dimensional RH equations and are \new{therefore} out of the scope of this work.

To calculate regular reflections, \new{CT-SD} uses the routine \rev{\texttt{shockObliqueReflectedTheta}} to compute the incident wave by specifying the wave angle $\beta$ (or the flow deflection $\theta$) and the pre-shock velocity $u_1$. This results in the calculation of the post-shock state (2), which serves as pre-shock state for the reflected wave (see sketch in Fig.~\ref{fig:validation_shocks_RR}\new{b}). If the incident shock is sufficiently strong, the transition to state (2) can result in significant thermochemical effects that may cause changes in the aerothermal properties compared to those of a calorically perfect fixed-composition gas. In this case, the values of $\mathcal{M}_2$ and $\theta$ may change accordingly. The reflected shock increases the gas pressure and temperature even further, and the properties in state (3) can be determined using the code routine employed for single oblique shocks imposing the counter deflection angle $\theta$ calculated for the incident shock. This guarantees that the streamlines in state (3) are  parallel to the reflecting surface. If there is no solution for the reflected shock (which occurs for sufficiently large values of $\theta$), irregular Mach reflections occur. As discussed above, these reflections involve non-uniform flow properties and non-steady solutions, and thus their computation is beyond the capabilities of \new{our code}.

As an illustrative example, Fig.~\ref{fig:validation_shocks_RR} represents the pressure-deflection shock polar diagrams for a regular shock reflection in atmospheric air at 30 km above sea level with an incident Mach number of $\mathcal{M}_1 = 20$ and a deflection angle of $\theta = 35^\circ$ under the same gas models \emph{ii)} and \emph{iii)} used in the oblique shock charts presented \new{previously}. The polar plot starting from state (2), obtained by increasing the incident wave angle from 0 to $\pi/2$, determines the solution of state (3) for the deflection angle $\theta = 35^\circ$. This, in turn, determines the reflected shock at the intersection of the second polar with the vertical axis ($\theta=0$). As can be seen, the dissociation and ionization effects are more pronounced in the reflected shock, as the accumulated temperature jump in both shocks amplifies the endothermicity \new{associated with} the \new{dissociation} reactions, resulting in substantially higher overall pressure ratios. Finally, the outcomes are compared with those acquired by Zhang et al.~\cite{zhang2022a} under the same flow conditions, revealing excellent agreement in all instances. In our calculations, the computation time was \rev{1.75 seconds} for a group of 28 species and 200 case studies. These values depend on the tolerance, which was set to $10^{-5}$ for the root-finding method.

\new{As previously discussed, for each ${\cal M}_1$, there exists a maximum value of $\beta$ (and, consequently, of $\theta$) beyond which regular reflection is impossible. This maximum value is determined using the \texttt{shockPolarLimitRR} method, which identifies the point where the polar diagram of the reflected shock becomes tangent to the $\theta = 0$ axis. Our code is able to compute $\beta_{\rm max}$ and $\theta_{\rm max}$ in cases involving high-temperature thermochemical effects. To determine this limit, we impose the condition $\theta_{3, \rm max} - \theta = 0$ and employ an iterative algorithm based on Broyden's method~\cite{dennis1996}, which makes use of the set of routines described above.}

\subsection{Planar gaseous detonations}
\label{sec:planar_detonations}

The thermochemical framework \new{used} to describe shock waves involving endothermic molecular transformations can be easily extended to account for exothermic reactions, as occurs in planar detonations. As with shock waves, the computation of detonations requires knowledge of the pre-shock state and the degree of overdrive that measures the contribution of the external supporting mechanism. However, unlike shock waves, detonations can be self-sustained, i.e., propagate without any external contribution exerting additional pressure from behind. This propagation mode, named after Chapman-Jouguet (CJ), involves the maximum possible expansion of the hot products. Therefore, the burnt-gas state is obtained by imposing the sonic condition in the post-wave flow $\mathcal{M}_2=1$. If the pressure behind the detonation wave is larger than what is anticipated by the CJ condition, which can only be achieved using an external forcing mechanism, the detonation is considered \textit{over-driven} and results in subsonic downstream conditions, with $\mathcal{M}_2<1$. Conversely, \textit{under-driven} detonations occur when the burnt gas is in a supersonic state with $\mathcal{M}_2>1$. \new{However, these types of waves are typically disregarded because the post-shock Mach waves within the non-equilibrium region diverge from the oblique shock wave, leading to acoustic decoupling from the wall and making wedge attachment impossible~\cite{Pratt1991}. Further studies also suggest that under-driven detonations may violate the second law of thermodynamics~\cite{Emanuel2004}. Nevertheless, since post-detonation equilibrium conditions are independent of the internal structure, under-driven detonations are not excluded by CT when calculating the complete spectrum of possibilities.}

The over-driven/under-driven solutions can be determined numerically for a defined upstream mixture and a given degree of overdrive $\eta = u_1/u_{\rm cj}$  (see functions {\texttt{detOverdriven}} and \rev{\texttt{detUnderdriven}}) by using the routines designed for CJ detonations (see \rev{\texttt{detCJ}}) and normal shocks (see \rev{\texttt{shockIncident}}). The \new{first routine} is \new{needed} to determine the minimum velocity $u_{1} = u_{\rm cj}$ (or $\eta=1$) required for a planar detonation to propagate, while the \new{second} is \new{used} to obtain the post-detonation state for a given degree of overdrive~$\eta$.

Careful selection of the initial guesses is \new{also} required to obtain the solutions for over-driven and under-driven detonations. For instance, $T_{2,\rm guess}$ and $p_{2,\rm guess}$ denote the estimated temperature and pressure after the detonation \new{front}, and should be anticipated considering the significant variations arising from the degree of overdrive and the type of propagation mode. To obtain the initial guesses for over-driven detonations, $p_{2,\rm guess}$ is computed using Eq.~\eqref{eq:RH}, assuming a constant $\gamma = \gamma_1 = \gamma_{2,\rm guess}$. This gives $p_{2,\rm guess} = p_1 (2 \gamma \mathcal{M}_1^2 - \gamma + 1) / (\gamma + 1)$. The temperature guess is based on the fact that, for sufficiently strong shocks, the kinetic energy downstream is much lower than upstream of the shock, $u_2^2/u_1^2\sim (\rho_1/\rho_2)^2 \ll 1$. This simplifies Eq.~\eqref{eq:RH} to $h_{2,\rm guess} = h_1 + u_1^2/2$, \new{thereby} enabling the computation of $T_{2,\rm guess}$ by solving the thermochemical equilibrium problem at specified enthalpy and pressure, $h_{2,\rm guess}$ and $p_{2,\rm guess}$. This approximation becomes more accurate with increasing degrees of overdrive, as the differences between $u_1$ and $u_2$ become more significant. This estimation method is the same as the one utilized in the incident normal shocks routine. For under-driven detonations, a reasonable initial guess can be obtained by considering the range of acceptable values for the mean post-shock density. By defining the dimensionless parameter $\zeta \in (0, 1)$ to measure how close $\rho_{2,\rm guess}$ is to the CJ state compared to the initial state, we can construct an initial guess for the density as follows: $\rho_{2,\rm guess} = [\zeta/\rho_{2,\rm cj} + (1-\zeta)/\rho_1]^{-1}$. The pressure and temperature values can then be determined using Eq.~\eqref{eq:RH} and the ideal gas EoS, respectively. It has been found that a value of $\zeta = 0.1$ is suitable for the set of Mach numbers tested.

\subsection{Oblique gaseous detonations}
Detonation waves can \new{also occur in} oblique configurations, \new{although they are much} less common \new{than} oblique shocks. However, \new{oblique detonations} are \new{critical} in Oblique Detonation Wave Engines (ODWE)~\cite{li1994, Kailasanath2000}, where the combustion process occurs \new{across} an oblique detonation \new{wave} that revolves around a cylindrical combustion chamber.

\begin{figure}[!ht]
    \centering
    \includegraphics[width=0.81\textwidth]{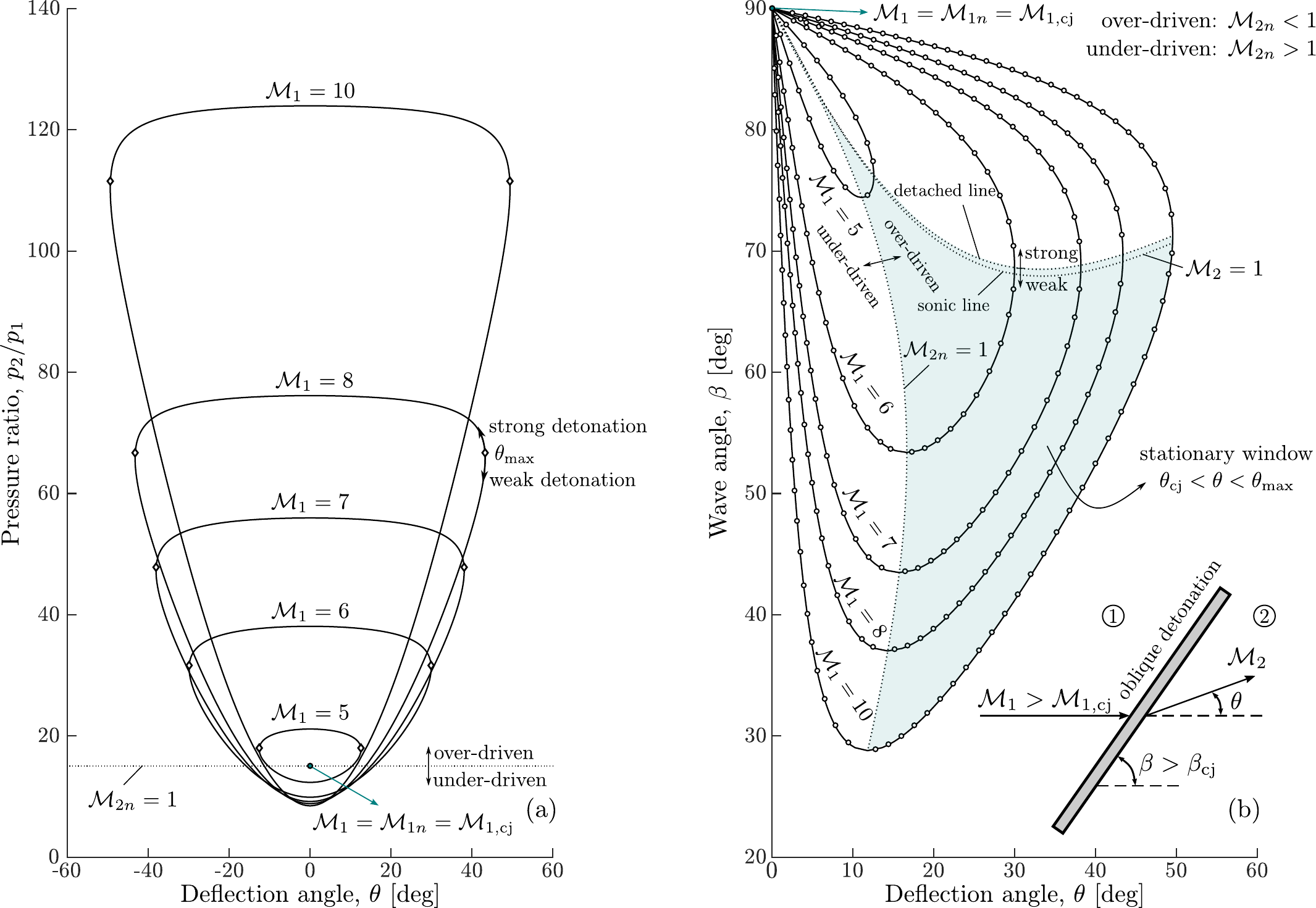}
    \caption{Pressure-deflection (a) and wave angle-deflection (b) detonation polar diagrams for a stoichiometric hydrogen (H$_2$) air (79\% N$_2$, 21\% O$_2$) mixture at pre-shock temperature $T_1 = 300$ K and pressure $p_1 = 1$ atm, and a range of pre-shock Mach numbers $\mathcal{M}_1$ between 5 and 10; solid line: results \new{from} CT considering a calorically imperfect gas with dissociation; circles: results from Zhang et al.~\cite{zhang2022a}.}
    \label{fig:det_polar}
\end{figure}

To compute oblique detonations, knowledge of the pre-shock state and the degree of overdrive caused by the supporting mechanism, typically a wedge deflecting a reactive supersonic stream and generating the oblique detonation, is required, just like in oblique shocks. Thus, given the temperature, pressure, composition, and pre-shock velocity, one can calculate the detonation polar diagrams with the particularity that now the exothermicity of the reaction increases the number of possible solutions, as occurs for planar detonations \cite{powers1992, zhang2022b}. Then, for a given detonation angle $\beta$ (or deflection angle $\theta$), two solutions for the burnt-gas state can be found, associated with under-driven ($\mathcal{M}_{2n}>1$) and over-driven ($\mathcal{M}_{2n}<1$) conditions (see routines \texttt{\rev{detonationObliqueBeta}} and \texttt{\rev{detonationObliqueTheta}}).

At the Chapman-Jouguet regime the two solutions merge into a single solution. This state is characterized by a sonic normal component of the downstream velocity vector ($\mathcal{M}_{2n}=1$). The corresponding values for the upstream Mach number and shock angle are $\mathcal{M}_{1}=\mathcal{M}_{1,\rm cj}$ and $\beta=\beta_{\rm cj}$. Both in the under-driven and over-driven cases, the RH-equations only produce real solutions if the values for the upstream Mach number and shock angle are \new{larger} than the corresponding values for the CJ condition, $\mathcal{M}_{1}\geq\mathcal{M}_{1,\rm cj}$ and $\beta\geq\beta_{\rm cj}$.  For oblique detonations with angles in the range $\beta_{\rm cj}<\beta<\pi/2$, a given shock angle corresponds to two different deflection angles, namely $\theta_{\rm over}$ and $\theta_{\rm under}$.

In the over-driven branch, the solution resembles that of an oblique shock.
The \new{condition} $\mathcal{M}_{2}=1$ is reached \new{just} below the maximum deflection angle separating the strong and weak solutions. Since the solution is multi-valued, the computation of the different branches requires accurate initial \new{guesses} regardless of the input parameter, be it the shock or the flow deflection angle (see Section~\ref{sec:planar_detonations}).

CT provides embedded functionalities for obtaining polar diagrams that characterize incident oblique detonations (see function \rev{\texttt{detPolar}}). These functionalities perform a direct computation of a set of cases (100 by default) by sweeping the range of possible solutions for both the under-driven and over-driven branches. \new{As an illustrative example, Fig.~\ref{fig:det_polar} shows the pressure-deflection (a) and wave angle-deflection (b) polar diagrams for detonations in stoichiometric hydrogen (H$_2$)-air (79\% N$_2$, 21\% O$_2$) mixtures at pre-shock temperature $T_1 = 300$~K and pressure $p_1 = 1$~atm, for a range of pre-shock Mach numbers $\mathcal{M}_1$ between 5 and 10.} The results \new{are} compared with \new{those of} Zhang et al.~\cite{zhang2022a}, which are found to be in remarkable agreement. The computation time was \rev{5.57} seconds for a set of 26 species and 1500 case studies\rev{ and} a tolerance of $10^{-5}$ for the root-finding method. For comparative purposes, the computation time required by Caltech's SD~Toolbox~\cite{Browne2008} with Cantera~\cite{Cantera}, which only provides the over-driven branch, was 101.77 seconds. This represents a \rev{$18\times$} speed-up factor for our code.

The blue-shaded area shown in Fig.~\ref{fig:det_polar} corresponds to weak over-driven conditions, which are the most likely to occur in oblique detonations. This region, bounded by $\theta_{\rm cj} < \theta < \theta_{\rm max}$, is of significant interest due to its applicability to the study of ODWE systems~\cite{zhuo2021, guo2021a}. For instance, recent research by Guo et al.~\cite{guo2021a} investigated the impact of pre-shock conditions on the stationary window for CH$_4$-air oblique detonations. The authors discovered that the limits  $\theta_{\rm cj}$ and $\theta_{\rm max}$ depend strongly on the pre-shock velocity and heat release associated with the mixture, while variations with the upstream temperature and pressure are not as prominent. Combustion Toolbox allows to perform these calculations easily, highlighting its relevance in carrying out preliminary studies before tackling more complex flow configurations.

\section{Rocket module}
\label{sec:ROCKET}

The calculation of the theoretical performance of rocket engines has drawn renewed attention in recent times \new{due to} the emergence of \new{new} private space companies such as Virgin Galactic, SpaceX, Blue Origin, Rocket Lab, \new{or} PLD Space, focused \new{on the development of} low-cost and reusable launch vehicles~\cite{PLD}. Despite the inherent complexities of these systems associated with the variety of physicochemical phenomena involved, a reasonably accurate estimation of engine performance can be achieved \new{with fairly simple thermochemical calculations}. \new{In particular}, since rocket engines typically operate at moderate pressures, the ideal gas assumption can be applied without the need for more complex equations of state. Additionally, the long residence times of the reacting gases in the combustion chamber compared to the chemical reaction times allow \new{for} further simplification of the calculations. This simplification allows for the utilization of thermochemical equilibrium tools such as CT.

The initial release of CT-ROCKET incorporates the mathematical description proposed in \cite{gordon1994}\rev{, implemented within the \texttt{RocketSolver} class}. \new{This method relies on several simplifying assumptions, including one-dimensional flow, uniform cross-sectional area, negligible flow velocity in the combustion chamber inlet, adiabatic combustion, isentropic expansion in the nozzle, homogeneous flow, the ideal gas law, and continuity of temperature and velocity} between gaseous and condensed species. Further details on the numerical implementation can be found in Ref.~\cite[Chapter 6]{gordon1994}.

\begin{figure}[h!]
    \centering
    \includegraphics[width=0.98\textwidth]{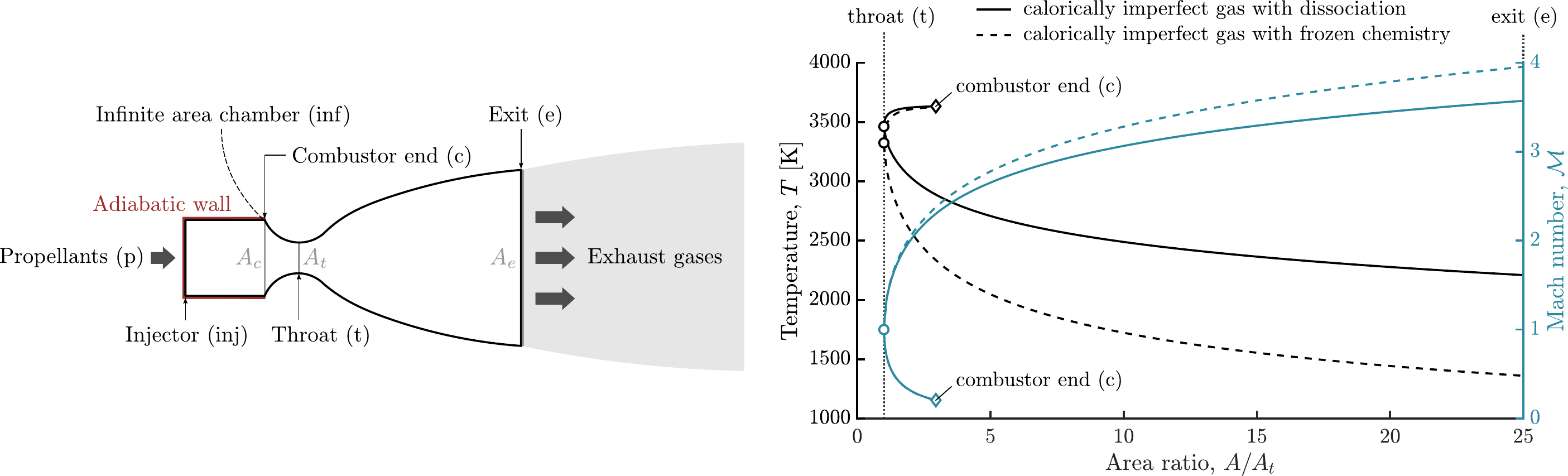}
    \caption{Sketch of the cross section of a finite area chamber (FAC) rocket engine. The dashed line represents the difference with an infinite area chamber (IAC), which is only included for the top region for clarity. Chemical transformations: (p-inj) and (p-inf) instant adiabatic combustion at constant pressure (HP); (inj-c) entropic process; (inf-t), (c-t), and (t-e) isentropic process at defined pressure (SP). Right: variation of the temperature (\protect\blackline) and Mach number (\protect\blueline) from the combustor end (c) to the exit (e) for a LOX/RP1 mixture with equivalence ratio $\phi = 1.5$ in a high-pressure combustion chamber $p_1 = 100$ atm considering calorically imperfect gas with dissociation (line) and calorically imperfect gas with frozen chemistry (dashed).}
    \label{fig:rocket_sketch}
\end{figure}

CT-ROCKET \new{leverages} the \new{ability of the} CT-EQUIL module to determine the gas composition within the rocket engine at various points of interest, such as the injector (inj), the combustion chamber outlet (c), the nozzle throat (t), and different points between (t) and (c/inf) where the hot gases are compressed (subsonic region) or between (t) and the nozzle outlet (e) where the hot gases \new{undergo the final expansion} (supersonic region), as illustrated in Fig.~\ref{fig:rocket_sketch} (left). The right panel shows the temperature and the Mach number of the fluid particles from the combustion chamber outlet (c) to the exit (e), passing through the throat (t) with $A_t=A_c/3$, for a LOX/RP1 mixture with equivalence ratio $\phi = 1.5$ (representing a 2.27 oxidizer/fuel weight ratio) in a high-pressure combustion chamber at $p_1 = 100$ atm. The area ratio is taken as the control variable, upon condition that thermochemical equilibrium is achieved at each position.

Additionally, the module calculates the thrust generated by the rocket engine. CT-ROCKET allows for calculations using either frozen chemistry or chemical equilibrium, accounting for combustion chambers with both finite (entropic process) and infinite (isentropic process) dimensions. The frozen chemistry and chemical equilibrium approaches provide an estimate of the performance limits of rocket engine nozzles, as demonstrated by Grossi et al.~\cite{grossi2023} through two-dimensional numerical simulations based on finite-rate kinetics. This feature enables the performance of parametric analyses to determine the optimal theoretical configuration for a given launch condition or to evaluate the environmental impact at various stages of the rocket vehicle. For instance, with the increasing number of space launches~\cite{jones2018,kokkinakis2022}, there has been a shift away from highly toxic fuels such as unsymmetrical dimethylhydrazine (UDMH) during the early phases of launch, towards so-called ``green" propellants like kerosene (RP1)~\cite{gohardani2014}. It is expected that the use of toxic fuels will be further restricted or even prohibited in the near future, thus driving the need to gain more experience with alternative ``green" propellants~\cite{dallas2020}. The CT-ROCKET module can \new{therefore} prove to be a valuable tool in this endeavor.

As previously discussed, this module also includes various routines to compute the state of the mixture at different points of the rocket engine. These states can be modeled using either an infinite area chamber (IAC) or a finite area chamber (FAC) \rev{implemented through the \texttt{rocketIAC} and \texttt{rocketFAC} methods, respectively. For example, in the FAC model,} an iterative procedure is employed to determine the mixture states at the chamber outlet (c) and at the throat~(t)\rev{. This is achieved by utilizing the IAC} model upon defining the initial fresh \rev{mixture's composition, temperature, and pressure, along with the area ratios} $A_c/A_t$ and $A_e/A_t$.

\begin{figure}[!htpb]
    \centering
    \includegraphics[width=0.98\textwidth]{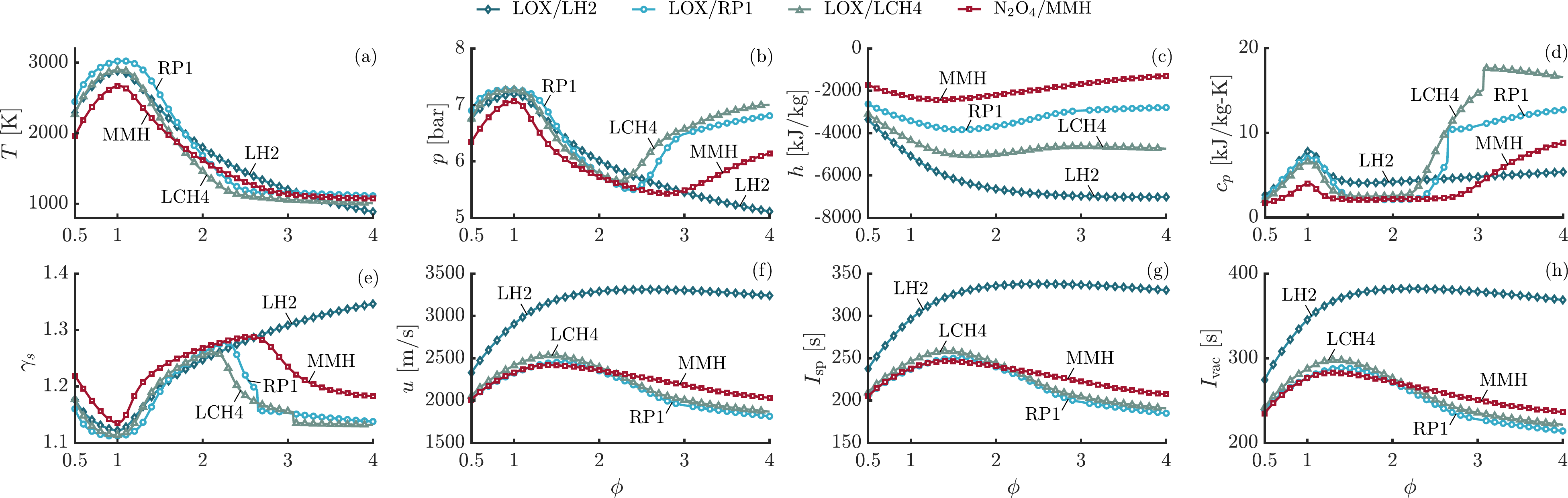}
    \caption{Thermodynamic properties at the nozzle exit of a rocket engine with aspect ratios $A_{\rm c}/A_{\rm t} = 2$ and $A_{\rm e}/A_{\rm t} = 3 $ for different liquid bi-propellant mixtures in a high-pressure combustion chamber, $p_1 = 100$ atm, with equivalence ratios $\phi \in [0.5, 4]$: temperature, $T$ (a), pressure, $p$ (b), enthalpy, $h$ (c), specific heat capacity at constant pressure, $c_p$ (d), adiabatic index, $\gamma_s$ (e), gas velocity, $u$ (f), specific impulse at sea level, $I_{\rm sp}$ (g), specific impulse in a vacuum, $I_{\rm vac}$ (h); solid line: results obtained with CT; symbols: results obtained with the NASA's CEA~\cite{gordon1994}: LOX/LH$_2$ (\footnotesize \setcolordiamonds{$\diamondsuit$}), LOX/RP1 (\raisebox{0.2mm}{\scriptsize\setcolorcircles{$\bigcirc$}}), LOX/LCH$_4$ (\raisebox{0.2mm}{\footnotesize\setcolortriangles{$\bigtriangleup$}}), N$_2$O$_4$/MMH (\raisebox{-0.1mm}{\footnotesize\setcolorsquares{$\square$}}).}
    \label{fig:rocket_results}
\end{figure}

Numerous validations were conducted using NASA's CEA code to ensure the reliability and robustness of CT-ROCKET. As an example, Fig.~\ref{fig:rocket_results} displays a range of thermodynamic properties computed at the nozzle exit (e). The geometrical aspect ratios defining the combustion chamber and the nozzle are $A_ {\rm c}/A_{\rm t} = 2$ and $A_{\rm s}/A_{\rm t} = 3$. Several reacting mixtures were utilized in the computations, including LOX/LH$_2$, LOX/RP1, LOX/LCH$_4$, and N$_2$O$_4$/MMH, the latter consisting of nitrogen tetroxide and monomethyl-hydrazine. The reactants were introduced in a combustion chamber where they reacted isobarically under high-pressure conditions, $p_1 = 100$ atm. The inlet temperature of the propellants was set to their respective boiling points, except for N$_2$O$_4$, which was evaluated at 300 K. The computations were performed over a wide range of equivalence ratios, $\phi \in [0.5, 4]$, to investigate the impact of the fuel-to-oxidizer ratio on the combustion process and the resulting reaction products.

As illustrated in \new{the different subplots of} Fig.~\ref{fig:rocket_results}, CT-ROCKET accurately predicts the properties of primary interest at the nozzle exit, including temperature ($T$) in (a), pressure ($p$) in (b), enthalpy ($h$) in (c), specific heat capacity at constant pressure ($c_p$) in (d), adiabatic index ($\gamma_s$) in (e), gas velocity ($u$) in (f), specific impulse at sea level ($I_{\rm sp}$) in (g), and specific impulse in a vacuum ($I_{\rm vac}$) in (h). The results demonstrate excellent agreement with the CEA code, with uniform convergence. However, the NASA code exhibited numerical instabilities for certain cases, such as LOX/RP1 at $\phi = 3$ and LOX/LCH4 at $\phi = 4$. The computation time for LOX/H2 was \rev{11.85} seconds for a set of 11 species and a total of 351 cases. The other mixtures were computed using 94 species, with an average computation time of \rev{30.88} seconds. It is noteworthy that the computation time per species is almost three times less for the latter cases. \rev{The performance of the CT-ROCKET module has improved by a factor of $1.9\times$ compared with the previous release~\cite{combustiontoolboxv1, cuadra2023_thesis}.}

\section{Graphic User Interface}
\label{sec:GUI}

\begin{figure}[!h]
    \centering
    \includegraphics[width=0.95\textwidth]{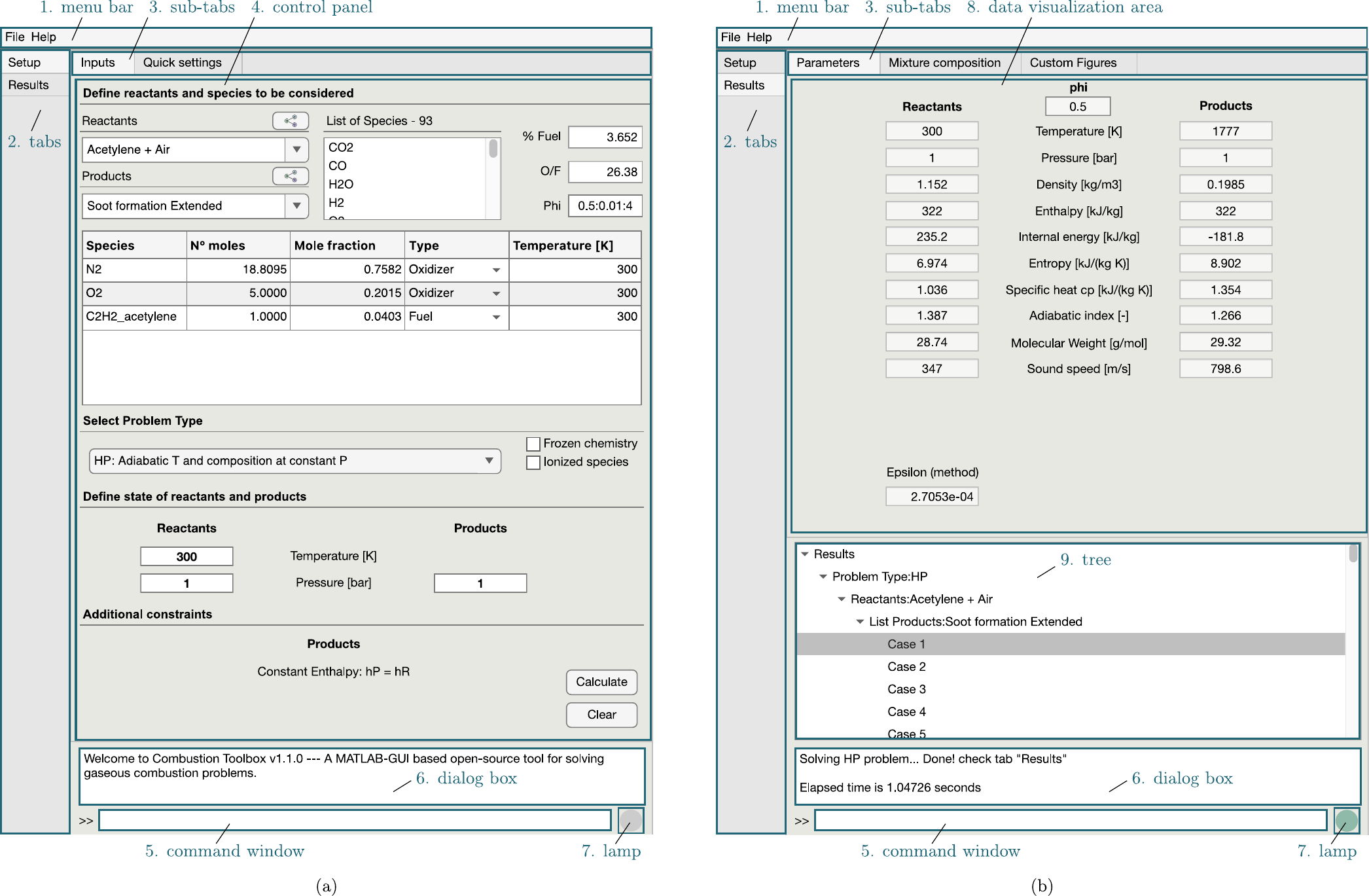}
    \caption{\new{Example of GUI configuration (a) and result post-processing (b) to reproduce the case of Fig.~\ref{fig:results_HP}. In particular, the thermodynamic properties correspond to the case selected in the tree object ($\phi = 0.5$).}}
    \label{fig:GUI_1}
\end{figure}

\begin{figure}[!h]
    \centering
    \includegraphics[width=0.95\textwidth]{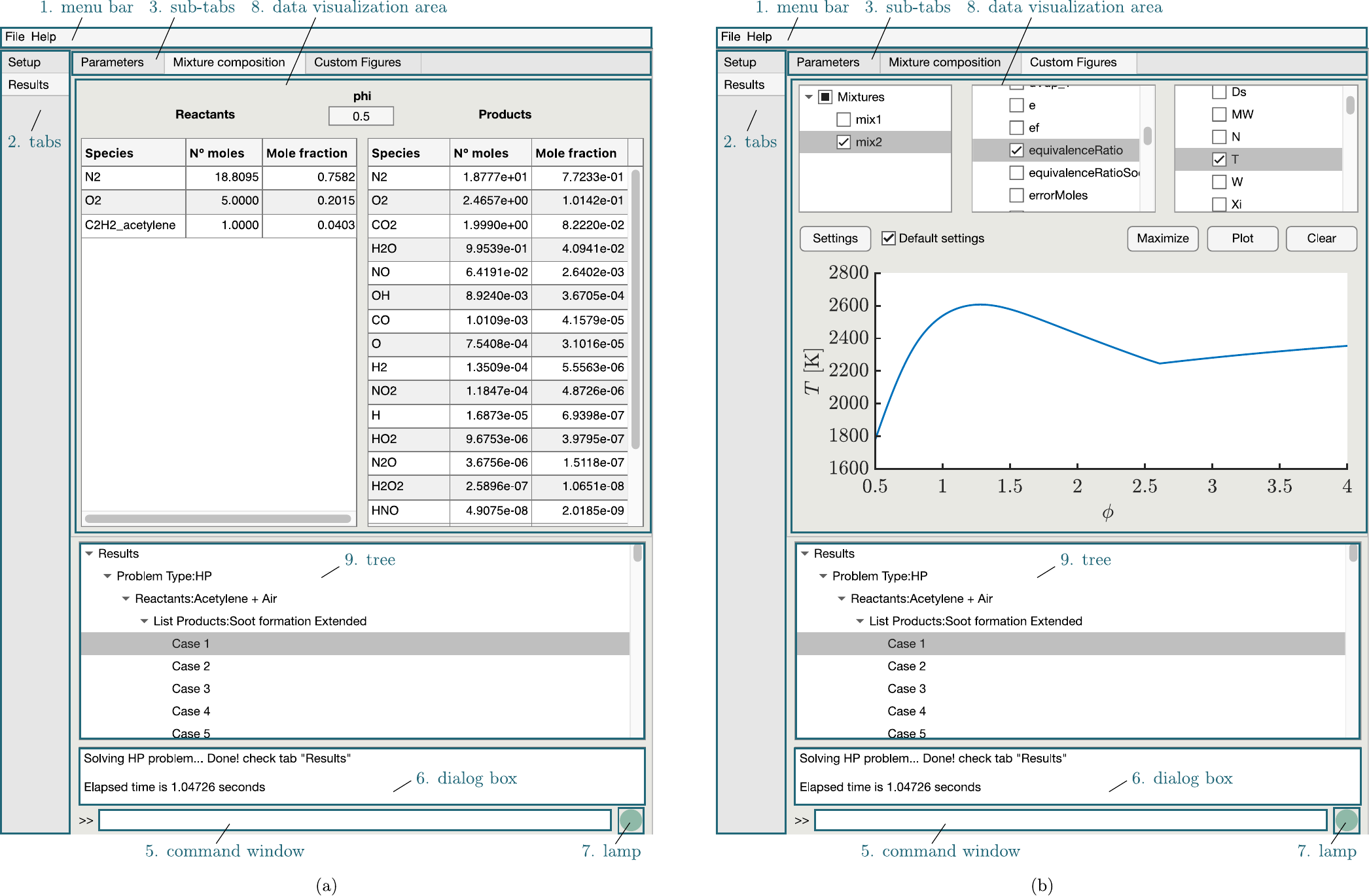}
    \caption{\new{Post-processing the results of Fig.~\ref{fig:results_HP} through the GUI: (a) chemical composition and (b) custom plots. The tab \textit{results} $\hookrightarrow$ \textit{mixture composition} shows the mixture composition for the case selected in the tree object ($\phi = 0.5$).}}
    \label{fig:GUI_34}
\end{figure}

This section presents a detailed overview of the Graphic User Interface (GUI) developed \new{for CT}. The GUI is intended to provide a user-friendly and intuitive interface for the visualization and analysis of data. Figures~\ref{fig:GUI_1} \new{and} \ref{fig:GUI_34} depict the fundamental elements of the GUI, described in detail below\new{:}

\begin{itemize}

    \item The \textit{menu bar} comprises predefined actions such as clear, save, snapshot, and check for updates, along with options to access online documentation, tutorials, examples, and license. It also provides access to additional tools, or add-ons, that can expand the GUI's capabilities. These add-ons include controls for numerical errors and visualization settings, species selection, as well as the ability to perform code validations and provide feedback to the development team (see CT documentation or its website).

    \item The interface is divided into two main \textit{tabs} (setup and results) and additional \textit{sub-tabs} designed to organize the content and prevent the user from feeling overwhelmed. The setup tab contains a control panel to configure the problem to be addressed. The results tab contains a data visualization area for post-processing all collected data.

    \item The \textit{control panel} is a crucial part of the GUI that enables users to configure the problem conditions. It provides a range of controls and options to adjust parameters such as the chemical species (reactants and products), the initial state (composition, temperature, and pressure), the type of problem to be solved, and other parameters for the setup of single-case or parametric studies.

    \item The \textit{command window} provides a command-line interface that allows users to interact with the GUI through a series of text commands. This feature is particularly useful for advanced users who prefer to work with code or scripts. Through this tool, users can input commands and execute scripts, while the \textit{dialog box} prompts the user for further input or confirmation before executing a command. The dialog box displays practical information like warnings, errors, and execution time, providing valuable feedback to the user.

    \item The \textit{lamp} component serves as a visual indicator of the analysis status. When the analysis is complete, the lamp emits a green light, while a yellow light indicates that the computations are still in progress. A red light indicates an error in the analysis.

    \item The \textit{data visualization area} displays the computed results in a visual format, such as plots or tables. It allows users to interact with and explore the data in a way that's intuitive and easy to understand.

    \item The \textit{tree} component collects all the data and exhibits the hierarchical organization of the obtained outcomes, which enables users to explore and access distinct aspects of each case.
    
    \item Additional features have been incorporated to improve the GUI's usability, including context menus and keyboard shortcuts. These functionalities enable users to perform intricate tasks with ease and speed.
\end{itemize}

For illustrative purposes, Figs.~\ref{fig:GUI_1} and \ref{fig:GUI_34} \new{show several} Combustion Toolbox GUI screenshots captured during \new{the} parametric study of adiabatic isobaric combustion of acetylene-air mixtures \new{reported as second validation test in Section~\ref{sec:equil_validations}. This case corresponds} to the data displayed in Fig.~\ref{fig:results_HP}. As observed in Fig.~\ref{fig:GUI_1}, the first step \new{involves} setting up the problem conditions, which includes \emph{i)} the initial mixture (composition, temperature, and pressure), that can be chosen from the predefined mixtures via the \textit{reactants} drop-down menu, or by manually adding the appropriate species name one-by-one to the same object; \emph{ii)} the problem configuration (adiabatic at constant pressure, etc.); \emph{iii)} the control parameter for a parametric or individual study, such as the equivalence ratio; and \emph{iv)} additional input parameters that may be required based on the type of problem.

If a parametric study is selected with the equivalence ratio as the control parameter, the post-processing step can be initiated as depicted in Fig.~\ref{fig:GUI_1}b. If the problem is well-posed, the \textit{lamp} object will turn green, meaning that the computations were \new{carried out} successfully, and a message will appears in the \textit{dialog box} of the GUI. Subsequently, the obtained dataset is re-structured in the background to fit into the \textit{tree} object for result post-processing. By selecting each solution of the \textit{tree} object, the GUI automatically updates the thermodynamic properties of the mixtures in the data visualization area (see Fig.~\ref{fig:GUI_1}). Additionally, in the \textit{results $\hookrightarrow$ \textit{custom figures}} tab, all mixture properties can be analyzed by plotting the results, as illustrated in Fig.~\ref{fig:GUI_34}. The datasets collected using the GUI can also be exported to a structured spreadsheet or \textit{.mat} file.

In brief, the GUI offers a broad range of tools for examining problems related to chemical equilibrium. The user-friendly design and intuitive features make it accessible to a variety of users, including those with a limited technical background. The GUI is an essential tool for researchers and practitioners who need to perform and analyse extensive parametric studies of the wide range of problems that can be addressed by the code. However, it is important to note that the GUI is intended to supplement traditional coding approaches instead of replacing them. Despite its ability to streamline tasks for non-experts, plain code actually exhibits greater versatility. The GUI itself is constructed upon an existing codebase, whose fundamental functions and calculations are still available for access and manipulation via the command line interface. In fact, proficient users interested in intricate analytical requirements may find that direct coding and execution is a more efficient and effective method than relying exclusively on the GUI. Thus, the GUI ought to be regarded as a tool that assists users with specific tasks rather than as a replacement for the potent and flexible nature of traditional coding.

\section{Conclusions}
\label{sec:conclusions}

In this work, we present the Combustion Toolbox (CT), an innovative open-source thermochemical code developed for the calculation of equilibrium states in gaseous reacting systems. While CT primarily focuses on combustion problems that may involve the formation of condensed-phase species, its abilities extend to other areas of interest, including the calculation of the atmospheric compositions of gaseous exoplanets, ablation processes, hypersonic shocks, and detonations. CT has been implemented in MATLAB and designed with \rev{an object-oriented} modular architecture, making it both user- and developer-friendly. Additionally, CT is equipped with an advanced Graphic User Interface that encapsulates the three modules and multiple built-in functions, providing users with a convenient operating experience.

At present, the three modules included in CT are CT-EQUIL, CT-SD, and CT-ROCKET. The first module, CT-EQUIL, is the kernel of the Combustion Toolbox and is responsible for solving the chemical composition of the system at equilibrium. This is achieved by minimizing the Gibbs/Helmholtz free using the Lagrange multiplier approach coupled with a multidimensional Newton-Raphson method. The second module, CT-SD, solves post-shock/detonation states for normal and oblique incident flows, including the computation of reflected waves. The third module, CT-ROCKET, is designed to determine the mixture composition at various points of interest within rocket engines, along with the calculation of the theoretical rocket performance.

The modules have been validated against existing state-of-the-art codes, including NASA's Chemical Equilibrium with Applications (CEA)~\cite{gordon1994}, Cantera~\cite{Cantera} within Caltech's Shock and Detonation Toolbox (SD-Toolbox)~\cite{Browne2008SDT, Browne2008}, and the newly developed Thermochemical Equilibrium Abundances (TEA) code~\cite{blecic2016}. All tests showed excellent agreement. As a matter of fact, CT exhibits superior computational performance in terms of cost and time, outperforming Caltech's SD-Toolbox and TEA by a significant margin. Additional validations can be accessed through the web at \url{https://combustion-toolbox-website.readthedocs.io}, which also provides further documentation and examples. The tool is actively maintained and can be accessed at \new{\url{https://github.com/CombustionToolbox/combustion_toolbox}}.

While the Combustion Toolbox has shown promising results, it is still an ongoing research project that requires additional development to enhance its capabilities. We aim to introduce additional functionalities in future versions of the code, such as the incorporation of non-ideal equations of state (currently under implementation), the analysis of multi-phase systems, a more accurate model for rocket propellant performance, \new{and} the extension of the database to include transport properties. We are also considering expanding the code to other well-known open-source programming languages, such as C++ and Python. The former is preferred due to its exceptional performance compared to MATLAB~\cite{andrews2012}, while the latter is \new{desired} due to its simplicity~\cite{fangohr2004}. An intermediate step will involve using MEX functions in the kernel of the code to combine C++ and MATLAB for calculating chemical equilibrium at defined temperature and pressure/volume, which is anticipated to substantially improve the speed of the code.

\section*{Acknowledgements}

The authors acknowledge funding from Comunidad de Madrid under the Multiannual Agreement H2SFE-CM-UC3M. C.H. also received support from project\new{s} TED2021-129446B-C41 \new{and PID2022-139082NB-C5} (MICINN/FEDER, UE). M.V. would like to express his lifetime gratitude for the invaluable and inspiring collaborations with Professors Amable Liñán and Antonio L. Sánchez over the past twenty-five years. Their seminal contributions, insight, motivation, and long term support played a pivotal role in initiating and promoting this research effort.

\section*{Declaration of Competing Interest}

The authors declare that they have no competing financial interests or personal relationships that could have appeared to influence the work reported in this paper.

 \newcommand{\noop}[1]{}

\end{document}